\RequirePackage{fix-cm}
\PassOptionsToPackage{dvipsnames}{xcolor}
\documentclass[epjc3,twocolumn,nopacs]{svjour3}
\usepackage{amsmath,amsfonts,amssymb,pifont,cite,orcidlink,physics,siunitx,verbatimbox,xspace}
\usepackage[utf8]{inputenc}
\usepackage[T1]{fontenc}
\usepackage{makecell}
\usepackage{physics}
\usepackage{booktabs}

\hypersetup{colorlinks=true,linkcolor=RedViolet,citecolor=RedViolet,filecolor=RedViolet,urlcolor=RedViolet}

\newcommand{\dmnlo}{{\sc DM@NLO}}

\newcommand{\cpp}{{\sc C++}}
\newcommand{\slhalib}{{\sc SLHALib-2.2}}
\newcommand{\LT}{{\sc LoopTools-2.16}}
\newcommand{\fortran}{{\sc Fortran}}
\newcommand{\pcm}{p_{\mathrm{cm}}}
\newcommand{\MOtwo}{{\sc MicrOMEGAs 2.4.1}}
\newcommand{\MO}{{\sc MicrOMEGAs}}
\newcommand{\DS}{{\sc DarkSUSY}}

\newcommand{\xmark}{\ding{55}}%
\newcommand{\cmark}{\ding{51}}%

\newcommand\DRbar{\ensuremath{\smash{\overline{\mathrm{DR}}}}}

\journalname{Eur.\ Phys.\ J.\ C}
\begin{document}

\title{Precision predictions for dark matter with \dmnlo\ in the MSSM}


\author{Julia Harz \thanksref{e2,addr2},
        Björn Herrmann\thanksref{e1,addr1},
        Michael Klasen \thanksref{e3,addr3},
        Karol Kova\v{r}\'\i{}k \thanksref{e4,addr3}
        \and
        Luca Paolo Wiggering\thanksref{e5,addr3} 
}

\thankstext{e2}{julia.harz@uni-mainz.de}
\thankstext{e1}{herrmann@lapth.cnrs.fr}
\thankstext{e3}{michael.klasen@uni-muenster.de}
\thankstext{e4}{karol.kovarik@uni-muenster.de}
\thankstext{e5}{luca.wiggering@uni-muenster.de}

\institute{PRISMA$^+$ Cluster of Excellence \& Mainz Institute for Theoretical Physics, FB 08 - Physics, Mathematics and Computer Science, Johannes Gutenberg University, 55099 Mainz, Germany  \label{addr2} \and
LAPTh, Université Savoie Mont Blanc, CNRS, 74000 Annecy, France \label{addr1} \and
Institut für Theoretische Physik, Universität Münster, Wilhelm-Klemm-Straße 9, 48149 Münster, Germany \label{addr3}
}

\date{Received: date / Accepted: date}

\maketitle

\vspace*{-8.0cm}
  \noindent {\small\texttt{MITP-23-084, MS-TP-23-52, LAPTH-062/23}}
\vspace*{7.0cm}

\begin{abstract}
 We present \dmnlo, a \textsc{Fortran 77} based program with a \textsc{C++} interface dedicated to precision calculations of dark matter (DM) (co)anni\-hila\-tion cross-sections and elastic dark matter-nucleon scattering amplitudes in the Minimal Supersymmetric (SUSY) Standard Model (MSSM) at next-to-leading order (NLO) in perturbative (SUSY) QCD. If the annihilating initial particles carry an electric or colour charge, the Sommerfeld enhanced cross section is included as well and can be matched to the NLO cross section. We review these calculations including technical details relevant for using the code. We illustrate their impact by applying \dmnlo\ to an example scenario in the constrained MSSM.
\end{abstract}

\section{Introduction}

The relic density of cold dark matter (CDM) in the cosmological standard ($\Lambda$CDM) model
\begin{align} 
    \Omega_{\rm CDM} h^2 ~=~ 0.120\pm0.001
    \label{Eq:PlanckCDM}
\end{align}
as 
determined from Planck data \cite{Planck:2018vyg} is one of the most stringent constraints on the nature and properties of DM. In the canonical freeze-out mechanism, today's DM abundance may be predicted for a given particle physics framework with a suitable DM candidate through all possible (co)annihilation processes of the thermal relic into Standard Model (SM) particles, and the most important channels are those with two particles in the final state \cite{Gondolo:1990dk,Jungman:1995df}.

Given the sub-percent accuracy of the Planck observation in Eq.\ \eqref{Eq:PlanckCDM}, matching the experimental and theoretical uncertainties clearly calls for the evaluation of the associated cross sections beyond the tree-level approximation. The impact of higher-order corrections to DM annihilation, both strong and electroweak, has been discussed within many well-motivated and intensely studied extensions of the SM such as the Minimal Supersymmetric Standard Model (MSSM) \cite{Freitas:2007sa, Herrmann:2007ku, Baro:2007em, Baro:2008bg, Baro:2009gn, Herrmann:2009wk, Herrmann:2009mp, Harz:2012fz, Harz:2014tma, Herrmann:2014kma, Harz:2014gaa, Schmiemann:2019czm, Branahl:2019yot, Klasen:2022ptb}, the Next-to-Minimal Supersymmetric Standard Model (NMSSM) \cite{Belanger:2016tqb, Belanger:2017rgu}, or the Inert Doublet Model \cite{Klasen:2013btp, Banerjee:2019luv, Banerjee:2021oxc}. The relic density can in addition potentially receive large corrections from non-perturbative effects like the Sommerfeld enhancement coming from the exchange of light mediators between the annihilating particles \cite{Iengo:2009ni, Beneke:2016ync, Harz:2017dlj} or the formation of bound-states \cite{vonHarling:2014kha, Harz:2018csl, Biondini:2018xor, Harz:2019rro}.

In addition to the increased precision, a further advantage of full loop calculations is that they allow for a systematical evaluation of the theoretical uncertainties from missing higher-order corrections through variations of the renormalisation scheme as well as the renormalisation scale \cite{Harz:2016dql, Branahl:2019yot}. 

The need for an increased theoretical precision extends to the calculation of indirect detection signals from present-day DM annihilation processes \cite{Hryczuk:2011vi, Hryczuk:2014hpa, Baumgart:2017nsr, Baumgart:2018yed} as well as to DM-nucleon interactions in the context of direct DM detection. Calculations of higher-order corrections to the corresponding scattering cross sections have been computed in many UV-complete models like the MSSM \cite{Drees:1992rr, Drees:1993bu, Hisano:2004pv, Berlin:2015njh, Klasen:2016qyz, Bisal:2023fgb,Bisal:2023iip}, the Inert Doublet Model \cite{Klasen:2013btp, Abe:2015rja}, simple Higgs-portal models \cite{Azevedo:2018exj, Ishiwata:2018sdi, Ghorbani:2018pjh, Abe:2018emu}, the Next-to-Minimal Two Higgs Doublet Model \cite{Glaus:2022rdc}, a vector DM model \cite{Glaus:2019itb}, the Singlet-Extended Two Higgs Doublet Model \cite{Biekotter:2022bxp}, but also in simplified fermionic DM models \cite{Berlin:2015ymu, Ertas:2019dew, Borschensky:2020olr}, or frameworks described through effective operators \cite{Haisch:2013uaa, Crivellin:2014qxa}.

In this context, we present in this paper the public release of the DM precision code \dmnlo \ ("Dark Matter at Next-to-Leading Order"), which allows to numerically calculate the total DM (co)annihilation as well as DM-nucleon scattering cross-sections at next-to-leading order (NLO) in the strong coupling constant for most (co)annihilation channels within the MSSM. In addition to the fixed-order one-loop calculation, the code includes several resummed corrections, such as the SUSY-QCD $\Delta m_b$ resummation and the Sommerfeld effect from the exchange of gluons or photons between the incoming particles.

Additional uncertainties not addressed within the context of \dmnlo\ stem from possible non-standard cosmologies \cite{Arbey:2008kv}, the neglect of multi-body final states \cite{Yaguna:2010hn}, uncertainties of the SM effective number of degrees of freedom \cite{Saikawa:2020swg}, differences in the numerical computation of the physical mass spectrum of the BSM model under consideration \cite{Allanach:2003jw}, thermal corrections \cite{Beneke:2014gla}, or the early-kinetic decoupling effect \cite{Binder:2017rgn, Aboubrahim:2023yag}. Moreover, on the side of the experimental analysis, the measured DM relic density relies on the the six-parameter cosmological concordance model. Including additional physical parameters \cite{Hamann:2006pf} or relaxing the assumption of a constant comoving DM density \cite{Bringmann:2018jpr} may increase the allowed range of the DM density by up to about 20\%. 

In the remainder of this manuscript, we begin by briefly summarising in Sec.\ \ref{sec:DMtheory} all relevant formulae for the calculation of the DM relic density in the standard freeze-out scenario and the spin-independent as well as spin-dependent elastic DM-nucleon scattering cross sections which are necessary to correctly interpret the \dmnlo\ output. The different components of a one-loop cross section as well as the Sommerfeld enhancement effect are described in Sec.\ \ref{sec:oneLoopTheory}. Sec.\ \ref{sec:howtodmnlo}, \ref{app:configuration}, and \ref{app:CLI} are dedicated to the installation and running of the program. An illustrative MSSM scenario is defined and analysed in Sec.\ \ref{sec:ill_examples}. We summarise this paper in Sec.\ \ref{sec:summary}.

\section{Dark matter theory}
\label{sec:DMtheory}

\subsection{Standard calculation of the relic density}

The standard procedure \cite{PhysRevLett.39.165, Gondolo:1990dk, Edsjo:1997bg} of calculating the freeze-out abundance of a single relic particle is usually based on the assumptions that {\it i)} the total (co)annihilation cross-section is governed by $2 \to n$ annihilation processes of DM into SM particles, {\it ii)} all dark sector particles are in kinetic equilibrium with the SM thermal bath at the photon temperature $T$ due to sufficiently large elastic scattering rates between both sectors, {\it iii)} these share the same chemical potential and {\it iv)} are highly non-relativistic so that in-medium as well as finite temperature effects are negligible. This means in particular that even for $n>2$, SM particles are assumed to obey a Maxwell-Boltzmann distribution. In this case, the evolution of the DM number density $n_\chi$ over time $t$ can be described with the single Boltzmann equation
\begin{equation}
    \dv{n_\chi}{t} + 3 H n_\chi ~=~ \langle \sigma_{\mathrm{ann} } v \rangle \Big( (n_\chi^{\mathrm{eq}})^2 - n_\chi^2 \Big) 
    \label{eq:boltzmann}
\end{equation}
with the equilibrium number density 
\begin{align}
    n_i^{\mathrm{eq}} ~=~ \frac{g_i T m_i^2}{(2\pi)^2} \, K_2(m_i/T)
\end{align}
of a particle species $i$ with mass $m_i$ and $g_i$ degrees of freedom for a vanishing chemical potential. The Hubble expansion rate $H$ is usually parameterised through the effective number of SM energy and entropy degrees of freedom. $K_i$ denote the modified Bessel functions of order $i$. Particle physics enters through the thermally averaged annihilation cross section 
\begin{equation}
    \langle \sigma_{\mathrm{ann}} v \rangle ~=~ \sum_{a,b} \langle \sigma_{a b\to X} v\rangle \frac{n_a^{\mathrm{eq}}}{n_\chi^{\mathrm{eq}}} \frac{n_b^{\mathrm{eq}}}{n_\chi^{\mathrm{eq}}}
    \label{eq:svavg}
\end{equation}
including all possible $2\to n$ (co)annihilation channels of dark sector particles into a set $X$ of $n$ SM particles. For a given initial-state configuration, the thermal average
\begin{align}\begin{split}
    \langle \sigma_{ab\to X} v\rangle 
    &= \frac{1}{2 \, T \,m_a^2 m_b^2 \, K_2(m_a/T) \, K_2(m_b/T)} \\ 
    &~~~~\times \! \int \! \dd{s} \sqrt{s} \, p^2_{\mathrm{cm}} \sigma_{ab\to X}(p_{\mathrm{cm}}) K_1(\sqrt{s}/T) 
\end{split}\end{align}
can be cast into a single integral over the centre-of-momentum energy $\sqrt{s} = \sqrt{m_a^2 + \pcm^2} + \sqrt{m_b^2 + \pcm^2}$ with $\pcm$ being the relative annihilation momentum. From the present-day number density $n_\chi^0$, the theoretically predicted dark matter relic density is obtained via 
\begin{align}
     \Omega_\chi ~=~ \frac{ m_\chi n^0_\chi}{\rho_c}
\end{align}
with $\rho_c$ being the critical density. The task of numerically integrating Eq.\ \eqref{eq:boltzmann} in conjunction with computing the thermal average in Eq.\ \eqref{eq:svavg} for a model dependent annihilation cross section can be performed with a high accuracy using adequate numerical codes, e.g., \MO \cite{Belanger:2018ccd}, \textsc{SuperIso Relic} \cite{Arbey:2009gu}, \textsc{MadDM} \cite{Ambrogi:2018jqj}, \DS \cite{Bringmann:2018lay} or \textsc{DRAKE} \cite{Binder:2021bmg}, which can even go beyond the assumption of kinetic equilibrium. 

\subsection{Calculation of the direct detection rate}

Results of direct dark matter detection experiments are usually presented as exclusion limits on the spin-dependent (SD) and spin-independent (SI) DM-nucleon scattering cross-sections, $\sigma_N^{\mathrm{SD}}$ and $\sigma_N^{\mathrm{SI}}$, as a function of the DM mass. However, as the typical energies in a direct detection experiment are much smaller than the heavy particle masses of the microscopic theory mediating the interaction between DM and the constituents of a nucleon, it is customary to perform the calculation in the language of an effective field theory \cite{Fan:2010gt,Hill:2014yka,Hill:2014yxa,Hisano:2015bma}, i.e. by integrating out those heavy mediators.

The spin-independent cross section
\begin{equation}
    \sigma_N^{\mathrm{SI}} = \frac{\mu_N^2}{\pi} |g_N^{\mathrm{SI}}|^2
\end{equation}
is then expressed through the SI effective DM coupling to nucleons $g_N^{\mathrm{SI}}$
with $\mu_N = m_N m_\chi/(m_N + m_\chi)$ being the reduced mass of the DM-nucleon system. The effective coupling is computed as
\begin{equation}
    g_N^{\mathrm{SI}} ~=~ \sum_q \bra{N}\bar{q} q\ket{N} \alpha^{\mathrm{SI}}_q \,,
\end{equation}
where the sum runs over all six quark flavors $q$ and $\alpha^{\mathrm{SI}}_q$ is the Wilson coefficient describing the SI interaction between quarks and the DM particle. The nuclear matrix element $\bra{N}\bar{q} q\ket{N}$ can be qualitatively understood as the probability of finding the quark $q$ inside the nucleon $N$ and is commonly expressed through the scalar nuclear form factors $f^N_{Tq}$ as
\begin{equation}
    \bra{N}m_q \bar{q} q\ket{N} ~=~ f^N_{Tq} \, m_N
\end{equation}
with the quark mass $m_q$ and the nucleon mass $m_N$. The scalar coefficients $f^N_{Tq}$ are determined from experiment and lattice QCD and are another source of theoretical uncertainties. To highlight the latter point, we show in Tab.\ \ref{tab:fTTable} the associated values that are hardcoded in \dmnlo\ and the two other DM packages \DS\ and \MO. The heavy quark form factors $f^N_{Tq}$ are obtained from those related to light quarks via the relation \cite{Shifman:1978zn}
\begin{equation}
    f_{Tc}^N ~=~ f_{Tb}^N ~=~ f_{Tt}^N ~=~ \frac{1}{27} \bigg(1 - \sum_{q=u,d,s} f_{Tq}^N \bigg) \,.
\end{equation}

The SD scattering cross section for DM on a single nucleon is given by
\begin{equation}
     \sigma_N^{\mathrm{SD}} ~=~ \frac{3 \mu_N^2}{\pi} \big| g_N^{\mathrm{SD}} \big|^2 \,,
\end{equation}
where the effective SD coupling $g_N^{\mathrm{SD}}$ between DM and nucleons reads
\begin{equation}
    g_N^{\mathrm{SD}} ~=~ \sum_{q=u,d,s} (\Delta q)_N \, \alpha^{\mathrm{SD}}_q
\end{equation}
with the SD Wilson coefficient $\alpha^{\mathrm{SD}}_q$ describing the DM-quark interaction. In contrast to the SI case, the sum runs only over the light quarks $u$, $d$ and $s$, as these carry the largest fraction of the nucleon spin which in turn is quantified through the axial-vector form factors $(\Delta q)_N$. The corresponding numerical values in \dmnlo\ are identified with those in \MO\ 5.3, given by
\begin{align}
\begin{split}
    (\Delta u)_p ~&=~ (\Delta d)_n ~=~  0.842 \,, \\
    (\Delta d)_p ~&=~ (\Delta u)_n ~=~ -0.427 \,, \\
    (\Delta s)_p ~&=~ (\Delta s)_n ~=~ -0.085 \,.
\end{split}
\end{align}

\begin{table}
    \renewcommand{\arraystretch}{1.3} \setlength{\tabcolsep}{1pt}\centering
    \label{tab:fTTable}
    \caption{Scalar nuclear form factors $f_{Tq}^N$ used in \dmnlo  \ based on Ref.\ \cite{Hoferichter:2015dsa}, \DS \ 6.4 based on Ref.\ \cite{Gasser:1990ce} and \MO \ 5.3 \cite{Belanger:2018ccd}.}
    \begin{tabular}{cccc}
        \hline
        Scalar coefficient & \dmnlo & \DS & \MO \\ 
        \hline 
        $f_{Tu}^p$  & 0.0208 & 0.023 &  0.0153 \\
        $f_{Tu}^n$ & 0.0189 & 0.019 & 0.0110 \\
        $f_{Td}^p$  & 0.0411 & 0.034 & 0.0191\\
        $f_{Td}^n$ & 0.0451 & 0.041 & 0.0273\\
        $f_{Ts}^p = f_{Ts}^n$  & 0.043 & 0.14 & 0.0447\\
        $f_{Tc}^p = f_{Tb}^p = f_{Tt}^p$ & 0.0663 & 0.0595 & 0.0682\\
        $f_{Tc}^n = f_{Tb}^n = f_{Tt}^n$ & 0.0661 & 0.0592 & 0.0679\\
        \hline
    \end{tabular} 
\end{table}

For a detailed discussion of direct detection in NLO SUSY-QCD, we refer the reader to Ref.\ \cite{Klasen:2016qyz}.

\section{Dark matter annihilation beyond tree-level}
\label{sec:oneLoopTheory}

At next-to-leading order (NLO), the tree-level DM (co)an\-nihilation cross-section is extended by the contribution  
\begin{equation}
    \Delta\sigma^{\mathrm{NLO}} ~=~ \int_n\dd{\sigma^{\mathrm{V}}} + \int_{n+1} \dd{\sigma^{\mathrm{R}}} \,,
\end{equation}
which contains virtual ($\dd{\sigma^{\mathrm{V}}}$) and real ($\dd{\sigma^{\mathrm{R}}}$) corrections, contributing at the same order in the coupling constant. In the present work, we focus on one-loop corrections in SUSY-QCD at order $\alpha_s$, including the emission of a real gluon. Note that, for certain couplings, we include the resummation of higher-order SUSY-QCD contributions as discussed in Refs.\ \cite{Herrmann:2007ku, Herrmann:2009wk, Herrmann:2009mp}.

\subsection{Renormalisation}
\label{sec:renschemes}

The virtual corrections are plagued by ultraviolet (UV) divergences whose removal requires a (numerically well behaved) renormalisation scheme, coming along with a suitable regularisation prescription. For the regularisation, the SUSY-preserving dimensional reduction (DR) scheme \cite{Siegel:1979wq, Signer:2008va} is employed, i.e.\ the relevant loop integrals are evaluated in $D = 4 - 2\epsilon$ space-time dimensions.

When it comes to the renormalisation of the MSSM, the squark masses have to be renormalised carefully since the stop and sbottom sectors have to be treated simultaneously due to the fact that the up- and down-type squarks share the common soft breaking parameter $M_{\tilde{q}}$ as a result of the $\operatorname{SU}(2)_L$ gauge symmetry. The squark mass matrix can be diagonalised,
\begin{align}
    U^{\tilde{q}} 
    \begin{pmatrix} 
    m^2_{LL} & m^2_{LR} \\ 
    m^2_{RL} & m^2_{RR}
    \end{pmatrix}
    (U^{\tilde{q}})^\dag  =  \begin{pmatrix} 
    m^2_{\tilde{q}_1} & 0 \\ 
    0 & m^2_{\tilde{q}_2}
    \end{pmatrix}
    \label{eq:squark_mixing}
\end{align}
with the two physical masses $m^2_{\tilde{q}_1}$ and $m^2_{\tilde{q}_2}$ being the eigenvalues of the non-diagonal mass matrix with the entries
\begin{align}
    &m^2_{LL} = M^2_{\tilde{Q}} + (I^{3 L}_q - e_q s_W^2) \cos2\beta m_Z^2 + m_q^2 \,, \\
    &m^2_{RR} = M^2_{\tilde{U},\tilde{D}} + e_q s_W^2 \cos2\beta m_Z^2 + m_q^2 \,,\\
    &m^2_{LR} = m^2_{RL} = m_q (A_q - \mu (\tan\beta)^{-2 I^{3L}_q}) \,.
\end{align}
Out of the eleven parameters $M_{\tilde{Q}}$, $M_{\tilde{U}}$, $M_{\tilde{D}}$, $A_t$, $A_b$, $m_{\tilde{t}_1}$, $m_{\tilde{t}_2}$, $m_{\tilde{b}_1}$, $m_{\tilde{b}_2}$, $\theta_{\tilde{t}}$ and $\theta_{\tilde{b}}$ only five are completely independent. As the renormalisation scheme should be applicable to all (co)annihilation channels with squarks in a leading role, we replace the soft SUSY-breaking masses $M_{\tilde{Q}}$, $M_{\tilde{U}}$, $M_{\tilde{D}}$ as input parameters by the physical on-shell masses $m_{\tilde{b}_1}$, $m_{\tilde{b}_2}$ and $m_{\tilde{t}_1}$. The three aforementioned soft parameters are then fixed through the requirement that Eq.\ \eqref{eq:squark_mixing} holds even at the one-loop order which, by inverting the corresponding eigenvalue equations, results in two possible solutions,
\begin{align}
m^2_{LL} &= \frac{m^2_{\tilde{q}_1}  + m^2_{\tilde{q}_2}}{2} \pm \frac{1}{2} \sqrt{(m_{\tilde{q}_1}- m_{\tilde{q}_2})^2 - 4 m_{LR}^4} \label{eq:mLL_sol} \,,\\
m^2_{RR} &= \frac{m^2_{\tilde{q}_1}  + m^2_{\tilde{q}_2}}{2} \mp \frac{1}{2}\sqrt{(m_{\tilde{q}_1}- m_{\tilde{q}_2})^2 - 4 m_{LR}^4} \,,\label{eq:mRR_sol}
\end{align}
for the diagonal entries of the mass matrix. Consequently, there are two possible values for $M_{\tilde{Q}}$, $M_{\tilde{U}}$, and $M_{\tilde{D}}$. However, not both of them may yield a numerically stable renormalisation scheme, the reason being that the diagonalisation may not correctly reproduce the mass of the lighter stop used as an input value, and, more importantly, the counterterm belonging to the heavier stop mass $\delta m_{\tilde{t}_2} \sim (U_{21}^{\tilde{q}} U_{12}^{\tilde{q}})^{-1}$ may become singular for vanishing off-diagonal elements of the squark mixing matrix. The same problem might occur in the counterterm related to the squark mixing angle $\delta \theta_{\tilde{q}} \sim (U_{11}^{\tilde{q}} U_{22}^{\tilde{q}} + U_{12}^{\tilde{q}} U_{21}^{\tilde{q}})^{-1}$. To avoid these issues, a scheme is defined as numerically stable if the following three conditions are fulfilled:
\begin{itemize}
    \setlength\itemsep{0.5em}  
    \item[$\bullet$] $|m^{\rm out}_{\tilde{t}_1} - m^{\rm in}_{\tilde{t}_1}|/m^{\rm in}_{\tilde{t}_1}< 10^{-5}$ \,,
    \item[$\bullet$] $|\mathrm{Re} \, U_{11}^{\tilde{q}} U_{21}^{\tilde{q}}| >10^{-4}$ \,,
    \item[$\bullet$] $|\mathrm{Re} \, (U_{11}^{\tilde{t}} U_{22}^{\tilde{t}} + U_{12}^{\tilde{t}} U_{21}^{\tilde{t}})| >10^{-4}$ \,.
\end{itemize}
Otherwise a scheme is declared as invalid (unstable). Given that both solutions are compatible, by default the solution is chosen where the dependent stop mass $m_{{\tilde{t}_2}}$ is closer to the corresponding physical value.

In a series of previous analyses \cite{Harz:2012fz, Herrmann:2014kma, Harz:2014tma, Schmiemann:2019czm, Klasen:2022ptb}, the following three renormalisation schemes, adapted to the present situation of DM (co)annihilation, have been introduced:
\begin{itemize}
    \item[0:] $m_b$, $m_t$, $m_{\tilde{f}}$, $\theta_{\tilde{f}}$, $A_f$ are all \DRbar\ parameters.
    \item[1:] $m_b$, $A_b$ and $A_t$ are \DRbar\ input parameters whereas $m_t$, $m_{\tilde{t}_1}$ $m_{\tilde{b}_1}$ and $m_{\tilde{b}_2}$ are on-shell (OS) masses. $\theta_{\tilde{t}}$, $\theta_{\tilde{b}}$ and $m_{{\tilde{t}_2}}$ are then dependent quantities.
    \item[2:] $m_t$, $m_b$, $A_b$ and $A_t$ are \DRbar\ input parameters and  $m_{\tilde{t}_1}$, $m_{\tilde{b}_1}$ and $m_{\tilde{b}_2}$ are OS masses. $\theta_{\tilde{t}}$, $\theta_{\tilde{b}}$ and $m_{{\tilde{t}_2}}$ are then dependent quantities. 
\end{itemize}
The hybrid on-shell/\DRbar\ scheme 1, which resembles the RS2 scheme presented in Ref.\ \cite{Heinemeyer:2010mm}, is the recommended option, whereas the other two schemes are well suited for the estimation of theoretical uncertainties from scheme variations. The integration of an automated selection of the best renormalisation scheme as, e.g., discussed in Ref.\ \cite{Heinemeyer:2023pcc} is left for a future update.

\subsection{Infrared treatment}

The real corrections on the other hand suffer from infrared (IR) divergent terms occurring in the soft or collinear phase-space regions which cancel against those singularities appearing in the one-loop diagrams of the virtual corrections. To make the analytic cancellation manifest and allow for the numerical integration over the real phase-space, we choose within \dmnlo \ the Catani-Seymour dipole subtraction method \cite{Catani:1996vz, Catani:2002hc} for massive initial-state particles \cite{Harz:2022ipe} over phase-space slicing methods \cite{Fabricius:1981sx, Giele:1991vf, Harris:2001sx}. This subtraction technique is based on the introduction of a local counterterm $\dd{\sigma^{\mathrm{A}}}$ that cancels the singularities in the real emission matrix element pointwise and is at the same time simple enough such that the integrals over the singular region can be performed analytically and the IR divergences appear as poles of the form $\varepsilon^{-1}$ and $\varepsilon^{-2}$. The whole procedure can be schematically captured in the equation 
\begin{align}
\begin{split}
    \Delta\sigma^{\mathrm{NLO}} ~&=~ \int_n \left[ \dd{\sigma^{\mathrm{V}}} + \int_1 \dd{\sigma^{\mathrm{A}}} \right]_{\varepsilon=0} \\ 
    &~~~~~~ + \int_{n+1} \Big[\dd{\sigma_{\varepsilon=0}^{\mathrm{R}}} - \dd{\sigma_{\varepsilon=0}^{\mathrm{A}}}\Big] 
\end{split}
\end{align}
and has the advantage over the slicing approach that no introduction of an arbitrary cutoff value on the energy or emission angle of the emitted gluon is required. This makes subtraction methods in general more numerically stable and therefore better suited for parameter space scans\footnote{For historical reasons, squark annihilation into electroweak final states is computed using the phase space slicing method. For stop annihilation into gluons and light quarks, the quarks of the first two generations are considered as effectively massless. Consequently, a consistent cancellation of infrared divergent terms requires the combination of the two processes $\tilde{t}_i \tilde{t}^\ast_j \to g g$ and $\tilde{t}_i \tilde{t}^\ast_j \to q\bar{q}$ at the loop level. To avoid double-counting, we therefore adopt the convention that the process with two gluons in the final state automatically includes the light quark contribution at both LO and NLO. For this reason, the code returns a zero cross section if the final state is set to a light quark-antiquark pair.}.

\subsection{Intermediate on-shell resonance subtraction}

Within the real emission contribution to the process $\tilde{\chi}^0_n \tilde{t}_i \to  b W^+$, the internal top propagator can become on-shell if the collisional energy $\sqrt{s}$ exceeds the top mass $m_t$. To cure the singularity, we follow the "Prospino scheme" defined in Refs.\ \cite{Beenakker:1996ch, Binoth:2011xi} and substitute the top propagator with the Breit-Wigner form, according to
\begin{equation}
     \frac{1}{p^2 - m^2_t} ~\to~ \frac{1}{p^2 - m^2_t + i m_t \Gamma_t} \,,
\end{equation}
in the resonant part $\mathcal{M}_{\rm r}$ of the total amplitude $\mathcal{M}_{\rm tot} = \mathcal{M}_{\rm r} + \mathcal{M}_{\rm nr}$, whereas the non-resonant piece  $\mathcal{M}_{\rm nr}$ remains unchanged. Since the corresponding process $\tilde{\chi}^0_n \tilde{t}_i \to t g$ is already accounted for in the calculation of the neutralino relic density, the contribution from the leading order on-shell production of a top with the subsequent decay into a bottom quark and a $W$-boson is removed locally through the replacement
\begin{equation}
     \big| \mathcal{M}_{\rm r} \big|^2 ~\to~ \big| \mathcal{M}_{\rm r} \big|^2 - \frac{m^2_t \Gamma^2_t}{(p^2 - m^2_t)^2 + m^2_t \Gamma^2_t}  \big| \mathcal{M}_{\rm r} \big|^2_{p_t^2=m_t^2}
\end{equation}
with the physical top width $\Gamma_t$. This procedure has the advantage that it retains the interference $\mathrm{Re}(\mathcal{M}^\ast_{\rm r} \mathcal{M}_{\rm nr})$ containing only one on-shell propagator and thus finite principal-value integrals. However, to stabilise the numerical integration, we use a small artificial top width $\Gamma_t = 10^{-3}\cdot m_t$ in the interference part instead of the physical width.

\subsection{Sommerfeld enhancement}

The Sommerfeld enhancement is an elementary quantum mechanical effect that increases (decreases) annihilation cross sections for small relative velocities in the presence of an attractive (repulsive) long-range potential affecting the incoming particles. From a field theory point of view, this effect is described by ladder diagrams involving the exchange of light mediators with some coupling $\lambda$ to the initial particles. More quantitatively, the Sommerfeld factor 
\begin{equation}
    S_0^{[\mathbf{R}]} ~=~ \big| \phi^{[\mathbf{R}]}(0) \big|^2
\end{equation}
is obtained as a solution $\phi(\vec{r})$ evaluated at the origin $\vec{r}=0$ of the stationary Schrödinger equation for the potential describing the interaction of the annihilating particles transforming under the representation $\mathbf{R}$ of the corresponding force carriers. For an $s$-wave dominated annihilation process, the Sommerfeld factor simply multiplies the perturbative tree-level cross section giving the Sommerfeld corrected cross section
\begin{equation}
    \sigma^{\mathrm{Som}} = \sum_{\mathbf{R}} S_0^{[\mathbf{R}]} \sigma^{\mathrm{Tree}}_{\mathbf{R}}\, , 
\end{equation}
where the sum runs over all irreducible representations contained in the decomposition of the initial particle pair. In this convention for the Sommerfeld factor, the free wave-function is normalised to one $|\phi_0^{[\mathbf{R}]}(0)|^2=1$ to ensure that $\sigma^{\mathrm{Som}} \to  \sigma^{\mathrm{Tree}}$ is fulfilled if the interaction governing the enhancement effect is turned off ($\lambda \to 0$). When combining the Sommerfeld effect with the full $\order{\lambda^2}$ correction, one has to be careful to not overcount the single mediator exchange contained in both calculations. Therefore, we make the decision to match both by removing the $\order{\lambda^2}$ contribution from the Sommerfeld factor. 

\section{Installing and running \dmnlo}
\label{sec:howtodmnlo}

\subsection{Installation}

The \dmnlo\ package is a high-energy physics program whose source code is written in \textsc{Fortran 77} with a \cpp \ interface similar to the precision code \textsc{Resummino} \cite{Fuks:2013vua,Fiaschi:2023tkq}. It is publicly available for download at \url{https://dmnlo.hepforge.org} and is licensed under the European Union Public Licence v1.1.

The code can be compiled with the GNU compiler collection (GCC) and \textsc{CMake} version 3.0 or higher. As external dependencies, the libraries \slhalib \ \cite{Hahn:2006nq} and \LT \ \cite{Hahn:1998yk} are required for reading particle spectra following the Supersymmetry Les Houches Accord 2 (SLHA 2) convention \cite{Skands:2003cj, Allanach:2008qq} and for evaluating one-loop integrals, respectively. The code ships directly with slightly modified versions of both libraries, as well as the \textsc{CUBA-1.1} \cite{Hahn:2004fe} library for performing multidimensional phase space integrals through the \textsc{VEGAS} Monte Carlo algorithm. 

Once downloaded, the code can easily be unpacked and installed by running the following commands in a shell:
\begin{verbatim}
    tar xvf DMNLO-X.Y.Z.tar
    cd DMNLO-X.Y.Z
    mkdir build
    cd build
    cmake .. [options]
    make
    make install
\end{verbatim}
The last command is optional and places the \dmnlo\ binary \texttt{dmnlo} as well as the static library \texttt{libdmnlo.a} in the top-level source directory, which is the setup we assume in the following. Otherwise, the executable can be found in \texttt{build/bin} and the library in \texttt{build/src}. To install the code, e.g.\ system wide, the installation directory can be set with the \texttt{cmake} option 
\begin{verbatim}
    -DCMAKE_INSTALL_PREFIX= 
\end{verbatim}
Compilers different from the default \textsc{C}, \cpp \ and \fortran \ compilers identified by \textsc{CMake} can be set with 
\begin{verbatim}
    -DCMAKE_<LANG>_COMPILER= 
\end{verbatim}
The path to an alternative \textsc{LoopTools} installation can be specified with \texttt{-DLOOPTOOLS=PATH} after setting \texttt{-DBUILD\_LOOPTOOLS} (default: \texttt{ON}) to \texttt{OFF} if libraries and headers are installed in the same folder, or through  \texttt{LOOPTOOLS\_INCLUDE\_DIR} and \texttt{LOOPTOOLS\_LIB\_DIR} if not.

After successful compilation, the local installation can be tested by running the commands
\begin{verbatim}
    ./dmnlo --help
    ./dmnlo input/DMNLO.in
\end{verbatim}
in a shell. The source files of the \cpp \ interface to \dmnlo \ are located in \texttt{src}, whereas the processes themselves implemented in \textsc{Fortran 77} are collected in the folder
\texttt{run\_dmnlo}. The name of each subfolder for every process supported by \dmnlo \ is summarised in Tab. \ref{tab:processes}, together with the key references documenting the corresponding calculational details. The directory \texttt{external} contains external dependencies like \textsc{LoopTools} or \textsc{SLHALib}. The folder \texttt{input/demo} provides for every process available in \dmnlo, sorted according the arXiv number of the corresponding publication, the associated example scenarios as SLHA 2 files as well as \textsc{Python 3} plotting routines that partially use \textsc{PySLHA} \cite{Buckley:2013jua} and allow to reproduce the most important cross section plots. 

\begin{table*}
    \renewcommand{\arraystretch}{1.3} 
    \setlength{\tabcolsep}{10pt}\centering
    \caption{List of (co)annihilation and elastic DM-nucleon scattering processes included in \dmnlo, given together with the location of the corresponding source code in \texttt{run\_dmnlo}, the references to the orignal publication and whether the Sommerfeld enhancement is included. Here, $\phi = \{h^0,H^0,A^0,H^\pm\}$, $V = \{Z^0,W^\pm,\gamma \}$, $\bar{V} = V \setminus \{ \gamma \}$, and $\ell$ ($\bar{\ell}$) can be any (anti)lepton. The indices can take the values $\{m,n\}=\{1,2,3,4\}$, $\{i,j\}=\{1,2\}$.}
    \label{tab:processes}
    \begin{tabular}{@{\extracolsep{\fill}}cccc@{}}
        \hline Process & Folder & References & Sommerfeld   \\ \hline
        $\tilde{\chi}^0_m \tilde{\chi}^0_n,\tilde{\chi}^\pm_i \tilde{\chi}^\pm_j,\tilde{\chi}^0_n \tilde{\chi}^\pm_i \to q\bar{q},q\bar{q}'$ & \texttt{ChiChi2QQ} & \cite{Herrmann:2007ku,Herrmann:2009wk,Herrmann:2009mp,Herrmann:2014kma} & \xmark \\
        $\tilde{\chi}^0_n \tilde{q}_i \to q'\phi,q' \bar{V},q'g$  with $q,q'  \in \{t,b\}$ &  \texttt{NeuQ2qx} & \cite{Harz:2012fz,Harz:2014tma,Harz:2022ipe}  &  \xmark  \\
        $\tilde{t}_1 \tilde{t}^\ast_1 \to VV,V\phi,\phi\phi,\ell\bar{\ell}$ & \texttt{QQ2xx} &\cite{Harz:2014gaa} & \cmark  \\
        $\tilde{q}_i\tilde{q}'_j \to q q'$ with $q,q' \in \{t,b\}$ & \texttt{stst2QQ} &\cite{Schmiemann:2019czm} & \cmark   \\
        $\tilde{\tau}_1 \tilde{\tau}^\ast_1 \to t \bar{t}$ & \texttt{staustau2QQ} & \cite{Branahl:2019yot} & \cmark \\
        $\tilde{t}_i \tilde{t}^\ast_j \to g g, q\bar{q}$ with $q \in \{u,d,c,s\}$  & \texttt{stsT2xx} & \cite{Klasen:2022ptb} & \cmark   \\ \hline 
        $ \tilde{\chi}^0_1 N \to \tilde{\chi}^0_1 N$ & \texttt{DD} & \cite{Klasen:2016qyz} & --\\
        \hline
    \end{tabular}
\end{table*}

\subsection{Running \dmnlo\ from the command line}

As indicated above, \dmnlo \ can be executed in a shell through the command
\begin{verbatim}
     ./dmnlo <dmnlo-input-file>
\end{verbatim}
where the mandatory argument \texttt{<dmnlo-input-file>} provides the path to a configuration file in plain text format specifying the process and corresponding input parameters. Details on all the available options in such an input configuration file are extensively documented in \ref{app:configuration}. One example input file delivered with the code is \texttt{input/DMNLO.in}. Alternatively, the parameter values defined in the input file can also be passed through the command line interface (CLI), which then supersedes the value included in the text file. In the following, all possible command line options are described. A concise summary is also provided in \ref{app:CLI}. The command line options follow the same naming convention as the variables in the configuration file, so that the transfer from the command line to the input file is straightforward.  We start with the general options that are valid for both the relic density as well as the direct detection module.

The path to the SLHA file containing the numerical values of masses, mixing angles and decay widths has to be defined with \texttt{-{}-slha}. The value of the renormalisation scale in \si{\giga\electronvolt} is fixed through the \texttt{-{}-muR} option whereas the renormalisation scheme must be set to one of the three schemes defined in Sec.~\ref{sec:renschemes}  with the option \texttt{-{}-renscheme}. The mixed \DRbar-OS scheme no. 1 is here the recommended option. 

The \texttt{-{}-choosesol} option defines the solution in the heavy quark sector, with 0 being the recommended option, where the solution is chosen such that the dependent stop mass $m_{{\tilde{t}_2}}$ is closest to the corresponding on-shell value from the SLHA file (see discussion above). The arguments 1 and 2 then correspond to the two solutions in Eqs. \eqref{eq:mLL_sol} and \eqref{eq:mRR_sol}, respectively. If \dmnlo \ is used from the command line and the renormalisation scheme fails, the code simply stops after issuing a warning.

Also included is a legacy option which can only be turned on through the CLI by passing the flag \texttt{-{}-legacy}. This mode defines the weak mixing angle $\theta_W$ and the $W$-mass as in the default MSSM model file in \MOtwo, i.e.\ $\sin\theta_W = 0.481$ and $m_W = \cos\theta_W m_Z$ with $m_Z$ being the on-shell $Z$-mass. We include this option since this definition was adopted in \dmnlo\ before the public release and allows to reproduce old results. Starting with \texttt{v1.0.0} the electroweak mixing angle is defined through the on-shell $Z$- and $W$-mass from the SLHA 2 file
\begin{equation}
    \sin^2\theta_W ~=~ 1 - \frac{m^2_W}{m^2_Z} \,.
\end{equation}
Note that the legacy option should only be used for the reproduction of previously published results. 

Lastly, the perturbative order of the calculation needs to be specified. This is only possible through the CLI via the arguments \texttt{-{}-lo} for LO-accurate predictions and \texttt{-{}-nlo} for NLO accuracy. For the calculation of (co)annihilation cross sections there are two more accuracy options. The flag \texttt{-{}-sommerfeld} returns the Sommerfeld enhancement alone, whereas \texttt{-{}-full} returns the NLO result matched to the Sommerfeld enhancement. Otherwise the highest order available is assumed. If no Sommerfeld enhancement is available, the \texttt{-{}-full} option returns just the NLO cross section.  

The initial and final particles of the (co)annihilation process are fixed according to the PDG numbering scheme \cite{ParticleDataGroup:2022pth}. The two options \texttt{-{}-particleA} and \texttt{-{}-particleB} fix the initial state, whereas the two produced SM particles must be referred to by setting \texttt{-{}-particle1} and \texttt{-{}-particle2}. The collisional energy $\sqrt{s}$ 
has to be defined with \texttt{-{}-pcm}, which is the center-of-mass momentum $\pcm$ of the incoming particles. The option \texttt{-{}-result} controls whether the output contains the total cross section $\sigma$ or the cross section times velocity $\sigma v$, both in units of $\si{\giga\electronvolt^{-2}}$, where the relative velocity is defined as $v = 2 \lambda^{1/2}(s,m_a^2,m_b^2)/s$ with $\lambda$ being the K\"all\'en function. 

The direct detection module is enabled through the \texttt{-{}-DD} option, which supersedes the specified (co)annihi\-lation settings. The output contains then the SI and SD scattering cross sections of DM on protons and neutrons in $\si{\centi\meter^2}$, respectively. The scalar nuclear form factors from Tab.\ \ref{tab:fTTable} can be found (and modified) in \texttt{DD/DD\_Init.F}. The \texttt{-{}-formfactor} option followed by an integer number (0 for \dmnlo, 1 for \DS\ and 2 for \MO) allows the user to select a set of values from Tab.\ \ref{tab:fTTable}. 

\subsection{The \dmnlo\ library}

To facilitate the usage of \dmnlo\ from within other codes, the static library \texttt{libdmnlo.a} provides the two functions 
\begin{verbatim}
   double cs_dmnlo(order, na, nb, n1, n2, 
      PcmIn, muR, &slha, rs, sol, &corrFlags)
   void dd_dmnlo(order, muR, &slha, rs, sol,
      ff, &cs)
\end{verbatim}
where the former returns the total (co)annihilation cross section and the latter writes the SI and SD DM-nucleon cross sections into the array \texttt{cs}. The renormalisation scale is set with \texttt{muR}, the SLHA 2 input file with \texttt{slha}, the renormalisation scheme through the flag \texttt{rs} and the associated solution for the three soft-breaking parameters with \texttt{rs}. The parameter \texttt{sol} corresponds to the \texttt{choosesol} option and \texttt{ff} in the argument set of the direct detection function to the \texttt{formfactor} option.

The integer \texttt{order} specifies the perturbative order of the calculation. Possible vales are 0 for the LO result, 1 for the NLO result. For the computation of the (co)annihilation cross section, two additional options are available for the $\texttt{order}$ parameter, namely 2 for the full result (including NLO calculation and Sommerfeld enhancement) and 3 for the Sommerfeld enhanced cross section alone (without including the NLO calculation).

The parameters $\texttt{na}$ and $\texttt{nb}$ are needed to fix the incoming particles through their respective PDG numbers, while $\texttt{n1}$ and $\texttt{n2}$ are meant to specify the two particles in the final state. The center-of-mass momentum is set through \texttt{PcmIn}. Finally, the integer array \texttt{corrFlags} allows to turn certain processes on and off, which may be useful if the corresponding contribution to the relic density is known to be negligible. 

The static library \texttt{libdmnlo.a} also provides the two functions 
\begin{verbatim}
   int canImprove_dmnlo(na, nb, n1, n2)
   int consistent_RS_dmnlo(rs, &slha, muR)
\end{verbatim}
The former allows to check whether a given process can be corrected with \dmnlo, while the latter verifies whether the particle spectrum contained in \texttt{slha} yields a stable renormalisation scheme.

Alternative to the manual decision what annihilation channels to include, the file \texttt{minimal\_example.cpp} located in  \texttt{external/micromegas\_5.3.41/MSSM} exemplifies the use of these functions with \MO \ in a way that only those channels contributing more than $\SI{2}{\percent}$ to the relic density are corrected. 

Before compiling the minimal example file through 
\begin{verbatim}
   make main=minimal_example.cpp
\end{verbatim}
the file \texttt{micromegas\_5.3.41/include/modelMakefile} has to be replaced with the associated modified version shipped with \dmnlo. This can be achieved by running
\begin{verbatim}
   tar xvfk micromegas_5.3.41.tgz
\end{verbatim}
in the \texttt{external/} directory where the \texttt{tar} option \texttt{-k} (or \texttt{-{}-keep-old-files}) ensures that our modified version of \texttt{modelMakefile} containing the paths to the required libraries according to the default installation of \dmnlo\ is retained. For different paths or an alternative \MO \ version, \texttt{modelMakefile} has to be adjusted accordingly by the user. After successful compilation, the \MO \ interface can be tested by running
\begin{verbatim}
    ./minimal_example Scenario.spc
\end{verbatim}
in a shell from the \texttt{MSSM} folder. For more details on the usage of the \texttt{corrFlags} argument, we refer to the explanation given in \texttt{minimal\_example.cpp}.  

\section{Illustrative example calculations}
\label{sec:ill_examples}

\begin{table}
\renewcommand{\arraystretch}{1.3} \setlength{\tabcolsep}{10pt}\centering
 \caption{Example scenario in the cMSSM with a positive Higgs supersymmetric mixing parameter $\mu$ where stop (co)annihliation is the dominant dark matter mechanism. All dimensionful quantities are in \si{\giga\electronvolt}.}
    \label{tab:cMSSM}
\begin{tabular}{@{\extracolsep{\fill}}cccc|cc@{}}
 \hline 
 $m_0$ & $m_{1/2}$ & $\tan\beta$ & $A_0$  & $m_{\tilde{\chi}^0_1}$ & $m_{\tilde{t}_1}$ \\ \hline
3000 & 1400 &  20  & 12000  & 606.3 & 648.3 \\\hline
\end{tabular}
\end{table}

\begin{figure*}
    \centering
    \includegraphics[width=0.49\textwidth]{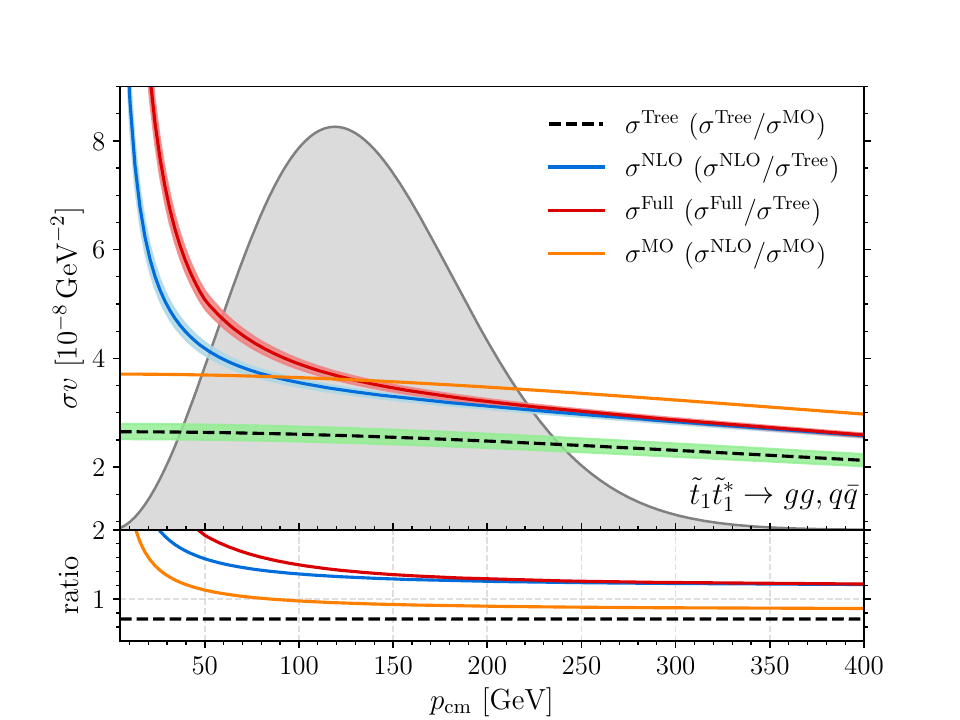}
    \includegraphics[width=0.49\textwidth]{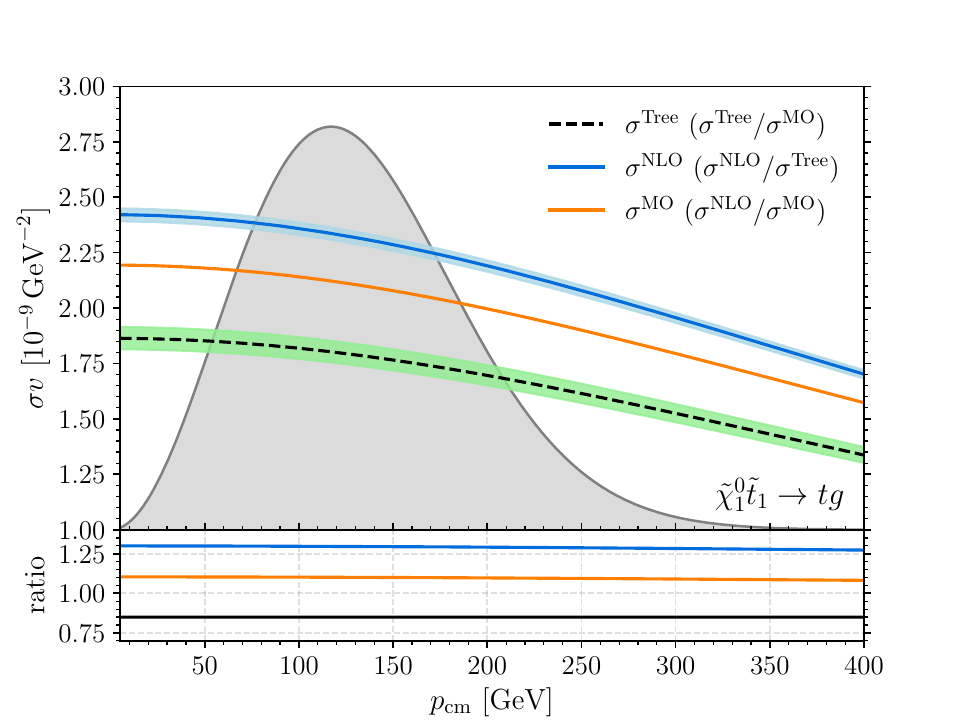}
    \includegraphics[width=0.49\textwidth]{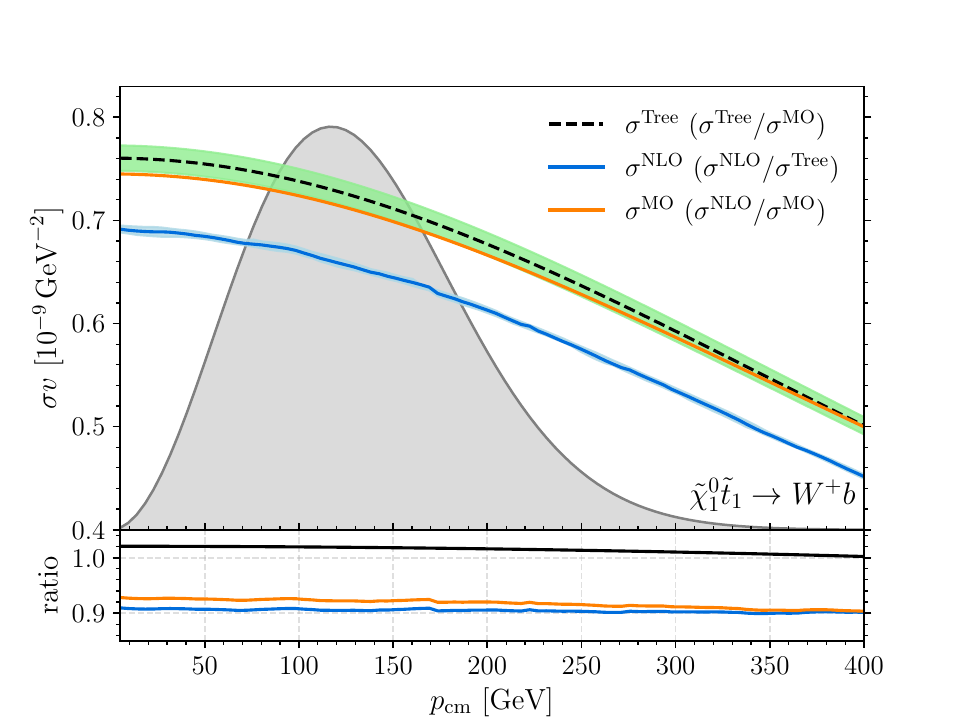}
    \includegraphics[width=0.49\textwidth]{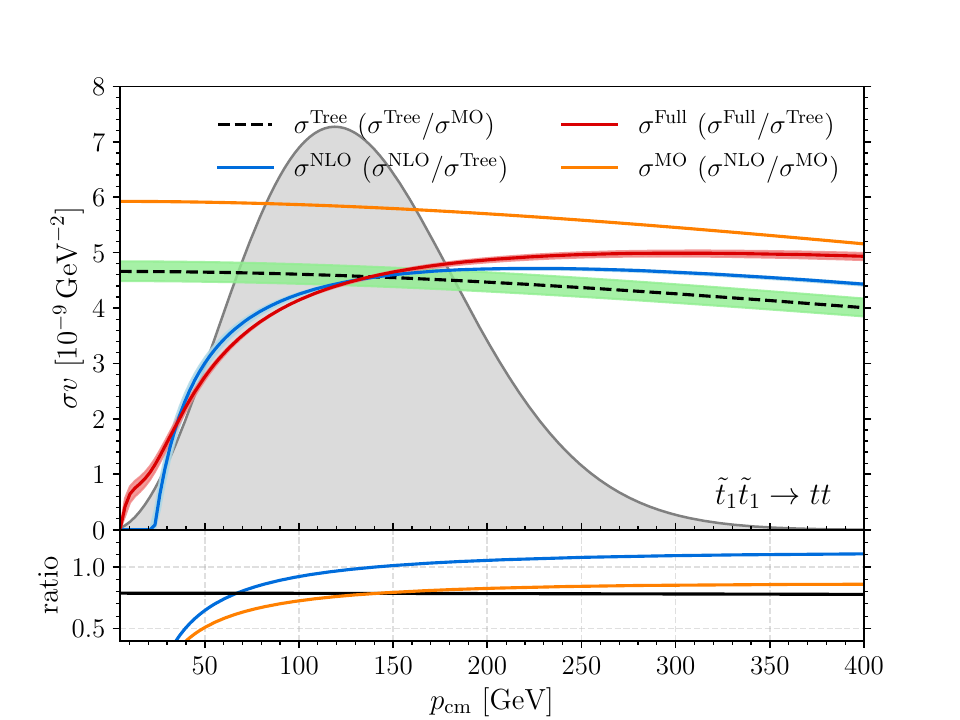}
    \includegraphics[width=0.49\textwidth]{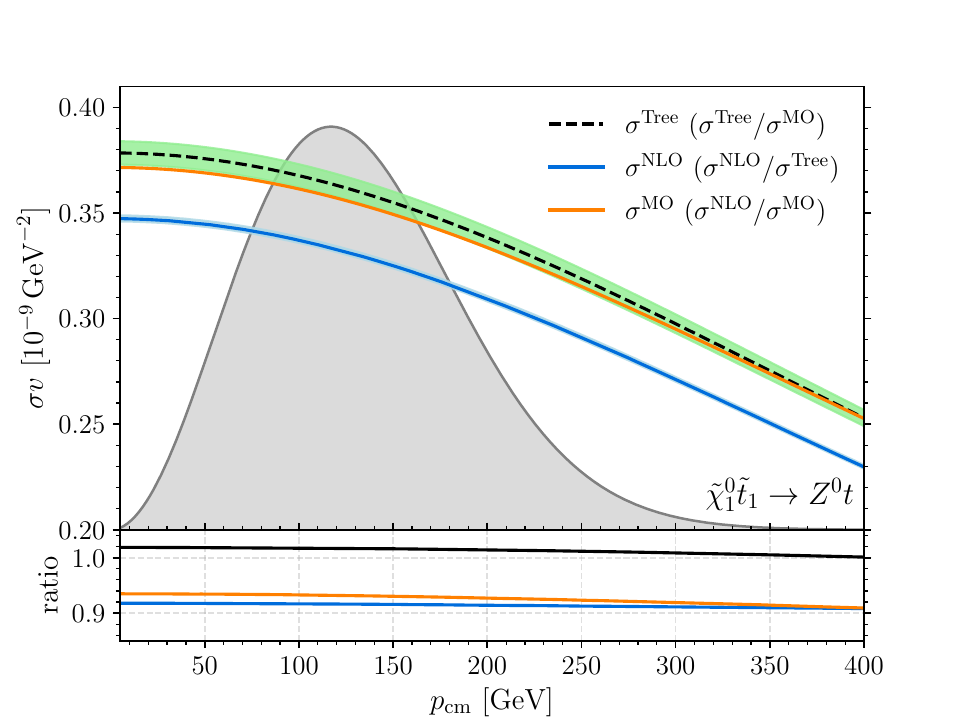}
    \includegraphics[width=0.49\textwidth]{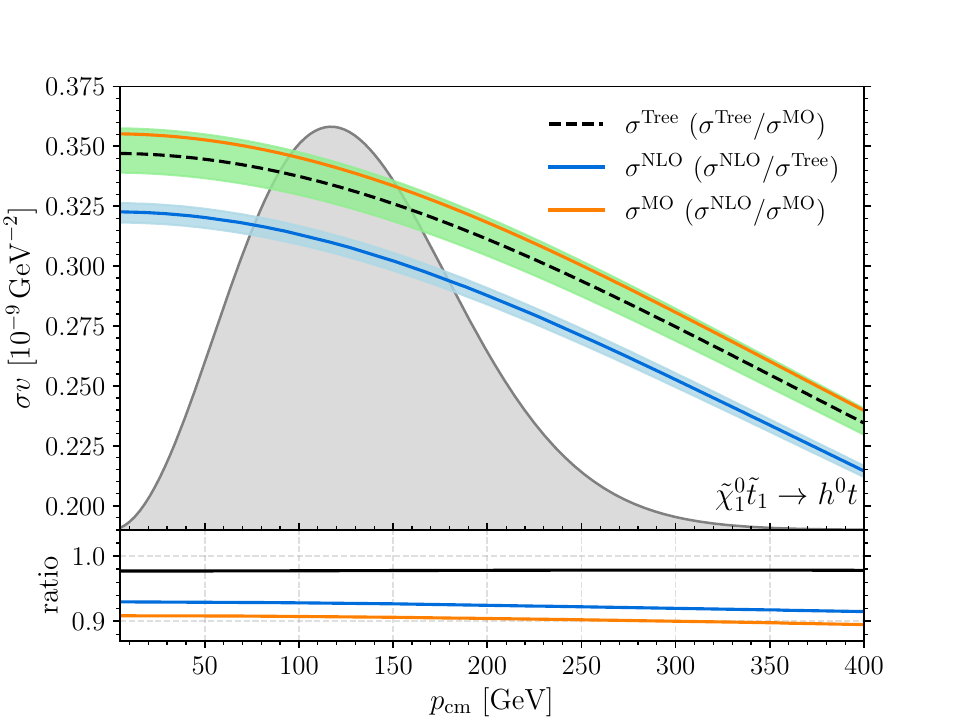}
    \caption{
    Tree-level (black dashed line), one-loop (blue solid line), full (red solid line) if present and \MO\ (orange solid line) cross sections for the dominant (co)annihilation channels shown in Tab. \ref{tab:channels} that can be corrected with \dmnlo \ including the corresponding uncertainties from variations of the renormalisation scale $\mu_R$ by a factor of two around the central scale as shaded bands. The upper part of each plot shows the absolute value of $\sigma v$ together with the thermal velocity distribution (in arbitrary units), whereas the lower part shows the corresponding relative shift (second item in the legend).}
    \label{fig:cMSSM_XSec}
\end{figure*}

To illustrate the usage of \dmnlo, we present example calculations in the constrained minimal supersymmetric extension of the Standard Model (cMSSM) which contains the simplifying assumption that the soft supersymmetry-breaking parameters unify at the gauge coupling unification scale of about $10^{16}$ GeV. This setup is entirely characterised through the universal scalar mass parameter $m_0$, the universal gaugino mass parameter $m_{1/2}$, the ratio of the vacuum expectation values of the neutral components of the two Higgs doublets $\tan\beta$, the universal trilinear coupling $A_0$, and the sign of the Higgs mixing parameter $\mu$. 

\begin{table}
\renewcommand{\arraystretch}{1.3} \setlength{\tabcolsep}{30pt}\centering
\caption{\label{tab:channels}
Dominant annihilation channels contributing to $\langle \sigma_{\rm ann}v \rangle$ for the cMSSM scenario in Tab.\ \ref{tab:cMSSM}. Further contributions below 2\% are omitted.}
\begin{tabular}{cc}
\hline Channel & Contribution \\ \hline
$\tilde{t}_1 \ \tilde{t}_1^\ast \rightarrow  g \ g$ & $\SI{36}{\percent}$   \\ 
$\tilde{\chi}^0_1 \ \tilde{t}_1\rightarrow t \ g $ & $\SI{29}{\percent}$  \\
$\tilde{\chi}^0_1 \ \tilde{t}_1\rightarrow W^+ \ b $ & $\SI{8}{\percent}$  \\
$\tilde{t}_1 \ \tilde{t}_1 \rightarrow  t \ t$  &   $\SI{6}{\percent}$  \\
$\tilde{t}_1 \ \tilde{t}_1^\ast \rightarrow \gamma \ g $ & $\SI{4}{\percent}$  \\
$\tilde{\chi}^0_1 \ \tilde{t}_1\rightarrow Z^0 \ t $ & $\SI{4}{\percent}$  \\
$\tilde{\chi}^0_1 \ \tilde{t}_1\rightarrow h^0 \ t $ & $\SI{4}{\percent}$  \\\hline
\texttt{DM@NLO} total  & $\SI{87}{\percent}$ \\\hline
\end{tabular}
\end{table}

In the following, we use, inspired by the recent search for non-excluded regions in the cMSSM parameter space \cite{Ellis:2022emx}, the parameter point given in Tab.\ \ref{tab:cMSSM}, where stop (co)annihilation is the dominant dark-matter annihilation mechanism\footnote{Note that numerical differences in the physical mass spectrum occur with respect to Ref.\ \cite{Ellis:2022emx} since \textsc{SPheno 4.0.5} is used as spectrum generator in this work whereas Ref.\ \cite{Ellis:2022emx} makes use of a private code. This is also the reason why $A_0 = 4 m_0$ is chosen in the example scenario in Tab.\ \ref{tab:cMSSM} versus $A_0 = 3 m_0$ in Ref.\ \cite{Ellis:2022emx}.}. The (co)annihi\-lation channels contributing most to $\langle \sigma_{\rm ann}v \rangle$ are listed in Tab.\ \ref{tab:channels} with stop-antistop annihilation into gluons having the largest relative contribution. For each channel that can be corrected with \dmnlo, we show in Fig. \ref{fig:cMSSM_XSec} our tree-level (black dashed line), the one-loop (blue solid line), and the full cross-section (red solid line), containing in addition to the NLO result the Sommerfeld enhancement, if available. For reference, the cross-section produced with the default \MO\ setup (orange solid line) is also shown. The grey shaded area depicts (in arbitrary units) the thermal velocity distribution, in order to demonstrate in which $p_{\rm cm}$ region the cross-section contributes to the total annihilation cross-section $\langle \sigma_{\rm ann}v \rangle$. In the lower part, the corresponding relative shifts of the different cross-section values (second item in the legend) are shown. Note that the difference between the \MO\ prediction and our tree-level result, which is particularly large for the $gg$ ($q\bar{q}$), $tg$ and $tt$ final states, is mainly due to a different choice of the renormalisation scale. We define the renormalisation scale through the tree-level stop masses as $\mu_R = \sqrt{m_{\tilde{t}_1} m_{\tilde{t}_2}}$, which for the particular scenario in Tab.\ \ref{tab:cMSSM} yields $\mu_R = 1368.2$ GeV, whereas \MO\ 5.3.41 uses the scale $Q = 2 m_{\tilde{\chi}^0_1}/3$ for the evaluation of the strong coupling stored in the global variable \texttt{GGscale}\footnote{This scale choice is different from \MO\ 2.4.1, where $Q = 2 m_{\tilde{\chi}^0_1}$ is used.}. We also show the uncertainties from variations of the renormalisation scale $\mu_R$ by a factor of two around the central scale as shaded bands.

\begin{figure*}
    \centering
    \includegraphics[width=.49\textwidth]{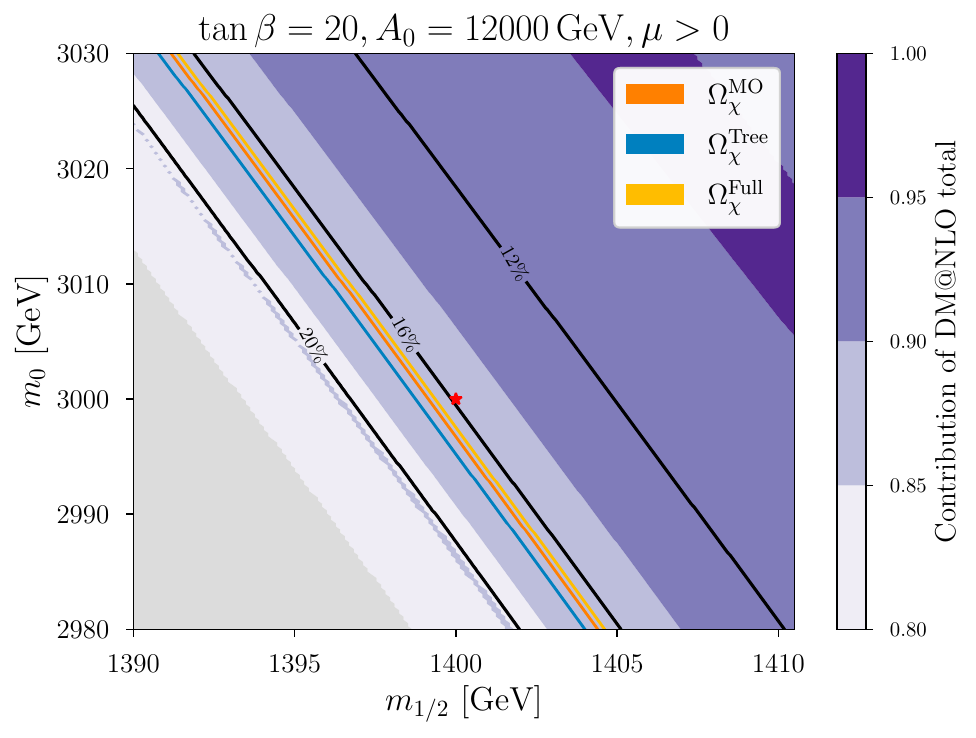}\quad
    \includegraphics[width=.49\textwidth]{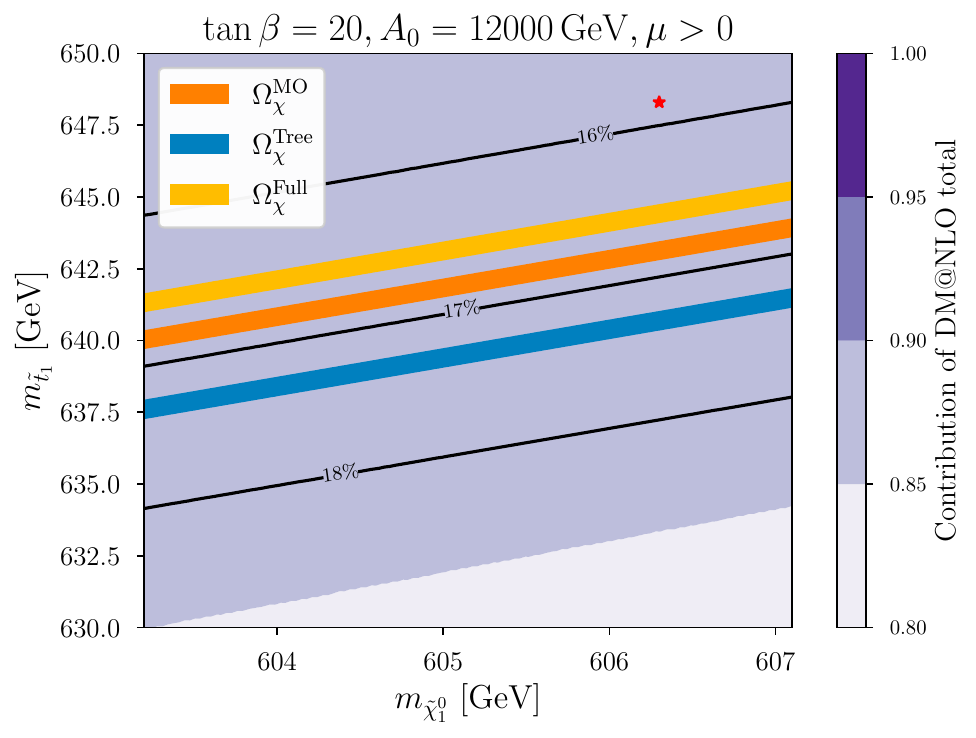}
    \caption{Bands compatible with the Planck measurement in Eq.\ \eqref{Eq:PlanckCDM} in the $m_{1/2}$-$m_0$ plane (left) and the plane 
    spanned by the associated physical masses of the lightest neutralino and the lightest stop (right) surrounding the example scenario from Tab.\ \ref{tab:cMSSM} shown in form of a red star. The three bands correspond to the \MO\ calculation (orange), our tree-level (blue) and our full corrections (yellow). The black solid lines indicate the relative change $(\Omega_\chi^{\rm Tree} - \Omega_\chi^{\rm Full})/\Omega_\chi^{\rm Tree}$ in the relic density compared to our tree-level result.}
    \label{fig:cMSSM_m0_m12}
\end{figure*}

In Fig.\ \ref{fig:cMSSM_m0_m12} we show the impact of our corrections on the neutralino relic density $\Omega_\chi h^2$ by performing a scan in the $m_{1/2}$-$m_0$ plane around the reference scenario of Tab.\ \ref{tab:cMSSM}, which is indicated by a red star. The orange band ($\Omega^{\rm MO}_\chi$) indicates the region consistent with the observed value $\Omega_{\rm CDM} h^2$ from Eq.\ \eqref{Eq:PlanckCDM} purely based on \MO\ and under the assumption that the lightest neutralino solely accounts for all of the observed DM, the blue band ($\Omega^{\rm Tree}_\chi$) corresponds to the prediction where the \dmnlo\ tree-level cross sections replace the CalcHEP result,
and the yellow band ($\Omega^{\rm Full}_\chi$) is based on our full calculation. 

The width of the three bands reflects the experimental $2\sigma$ uncertainty shown in Eq.\ \eqref{Eq:PlanckCDM}. One can observe a clear separation between all three bands everywhere across the shown $m_{1/2}$-$m_0$ plane. The black contour lines quantify the relative difference between our tree-level and our full calculation of the neutralino relic density. The increase amounts to roughly 16\% to 18\% in the regions consistent with the observed relic density $\Omega_{\rm CDM}$. Let us mention that \dmnlo\ allows to correct a large portion of the different contributions to the relic density which is between 85\% and 90\% in the relevant regions.

\begin{figure*}
    \centering
    \includegraphics[width=0.49\textwidth]{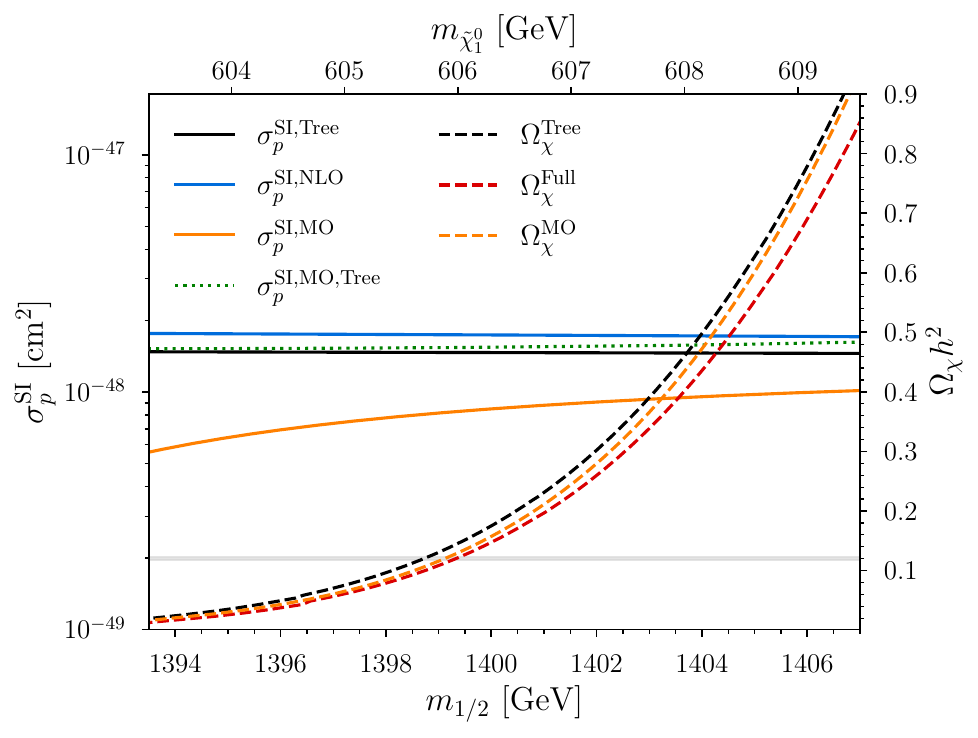}
    \includegraphics[width=0.49\textwidth]{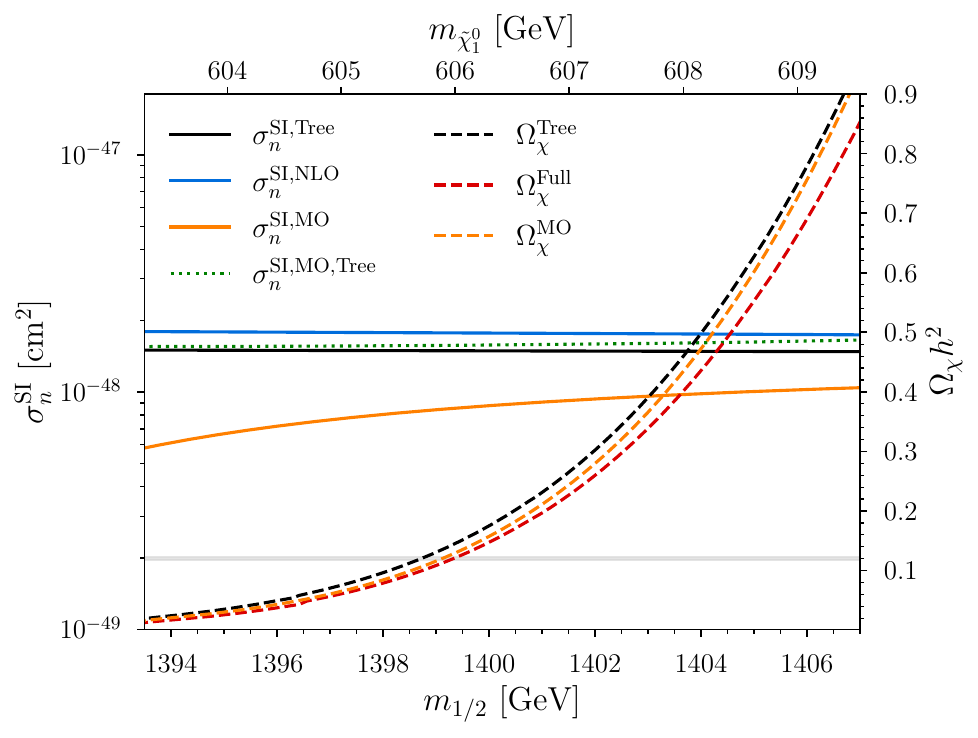}
    \includegraphics[width=0.49\textwidth]{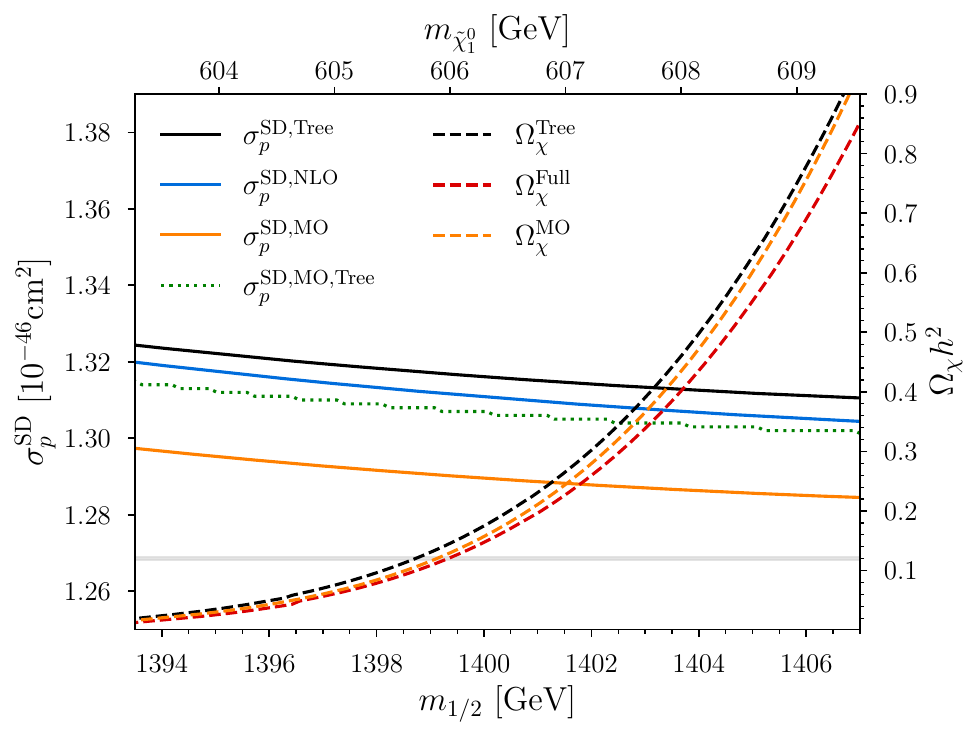}
    \includegraphics[width=0.49\textwidth]{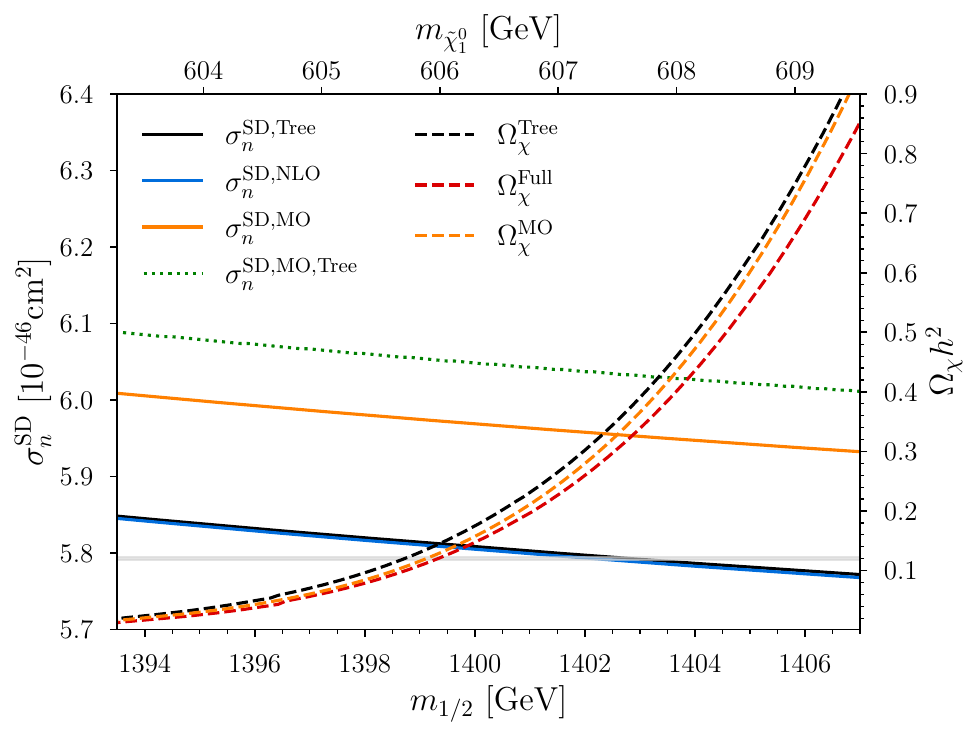}
    \caption{Spin-independent (top) and spin-dependent (bottom) neutralino-nucleon cross sections for protons (left) and neutrons (right) in the example scenario in Tab. \ref{tab:cMSSM} for different values of the universal gaugino mass parameter as well as the corresponding neutralino relic density obtained with \MO \ (MO), our tree-level calculation  (Tree) and with our full calculation (Full). The upper and lower limits imposed by Eq. \eqref{Eq:PlanckCDM} are indicated through the grey band.}
    \label{fig:Oh2_SI_SD}
\end{figure*}

Finally, we illustrate the application of \dmnlo\ to direct detection, i.e.\ we discuss the neutralino-nucleon cross-sections for different values of $m_{1/2}$ around the example scenario. In the upper panels of Fig.\ \ref{fig:Oh2_SI_SD}, the SI proton (left) and neutron (right) cross sections are shown, whereas the corresponding SD quantities are presented in the two lower panels. All quantities have been calculated with our \dmnlo\ code at tree level (black solid line), including the full $\order{\alpha_s}$ corrections to the dominant effective operators (blue solid line), \MO\ (orange solid line) and the corresponding analytic tree-level calculation\footnote{More precisely, the orange solid line corresponds to the \MO \, function \texttt{nucleonAmplitudes}, which includes additionally the box corrections calculated in Ref.\ \cite{Drees:1993bu}, while the green dotted line corresponds to the output of the function \texttt{MSSMDDtest(loop=0,\dots)} with the additional QCD corrections turned off \texttt{QCDcorrections=0}.}. 

We also show all three values of the resulting relic density ($\Omega^{\rm MO}_\chi$, $\Omega^{\rm Tree}_\chi$, $\Omega^{\rm Full}_\chi$) with the same colour coding as in Fig.\ \ref{fig:cMSSM_m0_m12}, as well as the Planck compatible value through a grey band. Note that the three curves increase as expected with the neutralino mass.

\section{Summary}
\label{sec:summary}

In this paper, we have presented the \dmnlo \ package, dedicated to precision calculation of dark matter (co)annihilation processes and direct detection in the MSSM. The program allows to compute total cross-sections at leading order and next-to-leading order in perturbative SUSY-QCD including the Sommerfeld enhancement effect. 

To illustrate the usage of the code and the impact of the higher-order corrections, various computations in a typical supersymmetric dark matter scenario were performed using \dmnlo. Apart from the benefit of having more precise predictions, we emphasise the possibility of estimating theory errors as a major advantage of using NLO cross sections and beyond. In addition, the general structure of the code is well-suited for the extension to non-supersymmetric models.

\begin{acknowledgements}
We would like to thank all colleagues who contributed to DM@NLO over the years, i.e. (in alphabetical order), J.\ Branahl, Q.\ Le Boulc'h, M.\ Meinecke, S.\ Schmiemann, P.\ Steppeler, and M.\ Younes Sassi. We would also like to thank A.~Pukhov for useful discussions throughout the development of \dmnlo, and A.~Puck Neuwirth for useful technical discussions on the build system. The research of MK and LPW was supported by the Deutsche Forschungsgemeinschaft (DFG, German Research Foundation) through the Research Training Group 2149 ``Strong and Weak Interactions - from Hadrons to Dark matter''. This work was also supported by \emph{Investissements d’avenir}, Labex ENIGMASS, contrat ANR-11-LABX-0012. JH acknowledges support from the Emmy Noether grant ``Baryogenesis, Dark Matter and Neutrinos: Comprehensive analyses and accurate methods in particle cosmology'' (HA 8555/1-1, Project No. 400234416) funded by the DFG and the Cluster of Excellence “Precision Physics, Fundamental Interactions, and Structure of Matter” (PRISMA$^+$ EXC 2118/1), funded by the DFG within the German Excellence Strategy (Project No.\ 390831469).
\end{acknowledgements}

\appendix

\section{The \dmnlo\ configuration file}
\label{app:configuration}

In this appendix, the options available in a \dmnlo \ configuration file are described. Such an input file consists out of a series of keywords that can be set to user defined values. These keywords are given by:
\begin{itemize} 
    \item \texttt{slha = <string>}: path to the SLHA input file defining the SUSY scenario to investigate (mass spectrum, mixing matrices, decay widths, {\it etc.}).\vspace{.1cm}
    \item \texttt{muR = <double>}: renormalisation scale $\mu_R$ in $\si{\giga\electronvolt}$.\vspace{.1cm}
    \item \texttt{renscheme = <int>}: renormalisation scheme according to the numbering scheme introduced in Sec. \ref{sec:renschemes}. The mixed \DRbar-OS scheme no. 1 is the recommended option. \vspace{.1cm}
    \item \texttt{choosesol = <int>}: solutions for $M_{\tilde{Q}}$, $M_{\tilde{U}}$, $M_{\tilde{D}}$ as explained in Sec. \ref{sec:renschemes}. \vspace{.1cm}
    \item \texttt{particleA = <int>} and \texttt{particleB = <int>}: PDG numbers of the first and second particle in the initial state.\vspace{.1cm}
    \item \texttt{particle1 = <int>} and \texttt{particle2 = <int>}: PDG numbers of the first and second particle in the final state.\vspace{.1cm}
    \item  \texttt{pcm = <double>}: centre-of-mass momentum in $\si{\giga\electronvolt}$.\vspace{.1cm}
    \item \texttt{result = <string>}: defines whether the output should contain the total cross $\sigma$ corresponding to the value \texttt{s} or the total cross section times relative velocity $\sigma v$ defined through the value \texttt{sv}. 
    \item \texttt{formfactor = <int>}: sets the scalar nuclear form factors $f^N_{Tq}$ to one of the sets of values in Tab. \ref{tab:fTTable}.
\end{itemize}

\section{Options available from the command line interface of \dmnlo }
\label{app:CLI}

The \dmnlo\ program can be run with several command line options by typing in a shell
\begin{verbatim}
    ./dmnlo <input-file> [options]
\end{verbatim}
The keyword \verb+<input-file>+ provides the path to a configuration file specifying the details of the computation to be achieved according to the standard defined in \ref{app:configuration}. The following options are allowed:

\begin{itemize} 
    \item \texttt{-{}-help}, prints a help message to the screen, indicating how to execute the code.\vspace{.1cm}
    \item \texttt{-{}-slha}, followed by a string sets the path to the SLHA 2 parameter file containing the numerical values of masses, mixing angles, decay widths, {\it etc}.\vspace{.1cm}
    \item \texttt{-{}-muR}, followed by a double-precision number sets the value of the renormalisation scale in $\si{\giga\electronvolt}$.\vspace{.1cm}
    \item \texttt{-{}-renscheme}, followed by an integer number sets the renormalisation scheme according to the numbering scheme introduced in Sec. \ref{sec:renschemes}. \vspace{.1cm}
    \item \texttt{-{}-choosesol}, followed by an integer number defines which solution to use for $M_{\tilde{Q}}$, $M_{\tilde{U}}$, $M_{\tilde{D}}$ as explained in Sec. \ref{sec:renschemes}. \vspace{.1cm}
    \item \texttt{-{}-legacy}, defines the weak mixing angle $\theta_W$ and the $W$-mass as in the default MSSM model file in \MOtwo. \vspace{.1cm}
    \item \texttt{-{}-lo}, returns the result at LO accuracy.\vspace{.1cm}
    \item \texttt{-{}-nlo}, returns the result at NLO accuracy.\vspace{.1cm}
    \item \texttt{-{}-sommerfeld}, returns the Sommerfeld enhanced cross section.\vspace{.1cm}
    \item \texttt{-{}-full}, returns the NLO result matched to the Sommerfeld enhancement if the latter is available. Otherwise the output is identical to \texttt{-{}-nlo}. \vspace{.1cm}
    \item \texttt{-{}-particleA} and \texttt{-{}-particleB}, followed by integer numbers defines the nature of the two initial-state particles through their PDG numbers.\vspace{.1cm}
    \item \texttt{-{}-particle1} and \texttt{-{}-particle2}, followed by integer numbers defines the nature of the two final-state particles through their PDG numbers.\vspace{.1cm}    
    \item \texttt{-{}-pcm}, followed by a double-precision number sets the centre-of-mass momentum $p_{\rm cm}$ in $\si{\giga\electronvolt}$.\vspace{.1cm}
    \item \texttt{-{}-result}, followed by a string corresponding to \texttt{s} for the total cross $\sigma$ or \texttt{sv} for the total cross section times the relative velocity $\sigma v$.\vspace{.1cm}
    \item \texttt{-{}-DD}, enables the direct detection module. This option supersedes (co)annihilation settings.
    \item \texttt{-{}-formfactor}, followed by an integer number ranging from zero to two sets the scalar nuclear form factors $f_{Tq}^N$ to one of the value sets shown in Tab. \ref{tab:fTTable} with zero for \dmnlo \ and two for \MO.
\end{itemize} 

\providecommand{\href}[2]{#2}\begingroup\raggedright\endgroup


\begin{thebibliography}{10}

\bibitem{Planck:2018vyg}
{\scshape Planck} collaboration, N.~Aghanim et~al., \emph{{Planck 2018 results.
  VI. Cosmological parameters}},
  \href{http://dx.doi.org/10.1051/0004-6361/201833910}{\emph{Astron.
  Astrophys.} {\bf 641} (2020) A6},
  [\href{http://arxiv.org/abs/1807.06209}{{\tt 1807.06209}}].

\bibitem{Gondolo:1990dk}
P.~Gondolo and G.~Gelmini, \emph{{Cosmic abundances of stable particles:
  Improved analysis}},
  \href{http://dx.doi.org/10.1016/0550-3213(91)90438-4}{\emph{Nucl. Phys. B}
  {\bf 360} (1991) 145--179}.

\bibitem{Jungman:1995df}
G.~Jungman, M.~Kamionkowski and K.~Griest, \emph{{Supersymmetric dark matter}},
  \href{http://dx.doi.org/10.1016/0370-1573(95)00058-5}{\emph{Phys. Rept.} {\bf
  267} (1996) 195--373}, [\href{http://arxiv.org/abs/hep-ph/9506380}{{\tt
  hep-ph/9506380}}].

\bibitem{Freitas:2007sa}
A.~Freitas, \emph{{Radiative corrections to co-annihilation processes}},
  \href{http://dx.doi.org/10.1016/j.physletb.2007.07.019}{\emph{Phys. Lett. B}
  {\bf 652} (2007) 280--284}, [\href{http://arxiv.org/abs/0705.4027}{{\tt
  0705.4027}}].

\bibitem{Herrmann:2007ku}
B.~Herrmann and M.~Klasen, \emph{{SUSY-QCD Corrections to Dark Matter
  Annihilation in the Higgs Funnel}},
  \href{http://dx.doi.org/10.1103/PhysRevD.76.117704}{\emph{Phys. Rev. D} {\bf
  76} (2007) 117704}, [\href{http://arxiv.org/abs/0709.0043}{{\tt 0709.0043}}].

\bibitem{Baro:2007em}
N.~Baro, F.~Boudjema and A.~Semenov, \emph{{Full one-loop corrections to the
  relic density in the MSSM: A Few examples}},
  \href{http://dx.doi.org/10.1016/j.physletb.2008.01.031}{\emph{Phys. Lett. B}
  {\bf 660} (2008) 550--560}, [\href{http://arxiv.org/abs/0710.1821}{{\tt
  0710.1821}}].

\bibitem{Baro:2008bg}
N.~Baro, F.~Boudjema and A.~Semenov, \emph{{Automatised full one-loop
  renormalisation of the MSSM. I. The Higgs sector, the issue of tan(beta) and
  gauge invariance}},
  \href{http://dx.doi.org/10.1103/PhysRevD.78.115003}{\emph{Phys. Rev. D} {\bf
  78} (2008) 115003}, [\href{http://arxiv.org/abs/0807.4668}{{\tt 0807.4668}}].

\bibitem{Baro:2009gn}
N.~Baro and F.~Boudjema, \emph{{Automatised full one-loop renormalisation of
  the MSSM II: The chargino-neutralino sector, the sfermion sector and some
  applications}},
  \href{http://dx.doi.org/10.1103/PhysRevD.80.076010}{\emph{Phys. Rev. D} {\bf
  80} (2009) 076010}, [\href{http://arxiv.org/abs/0906.1665}{{\tt 0906.1665}}].

\bibitem{Herrmann:2009wk}
B.~Herrmann, M.~Klasen and K.~Kovarik, \emph{{Neutralino Annihilation into
  Massive Quarks with SUSY-QCD Corrections}},
  \href{http://dx.doi.org/10.1103/PhysRevD.79.061701}{\emph{Phys. Rev. D} {\bf
  79} (2009) 061701}, [\href{http://arxiv.org/abs/0901.0481}{{\tt 0901.0481}}].

\bibitem{Herrmann:2009mp}
B.~Herrmann, M.~Klasen and K.~Kovarik, \emph{{SUSY-QCD effects on neutralino
  dark matter annihilation beyond scalar or gaugino mass unification}},
  \href{http://dx.doi.org/10.1103/PhysRevD.80.085025}{\emph{Phys. Rev. D} {\bf
  80} (2009) 085025}, [\href{http://arxiv.org/abs/0907.0030}{{\tt 0907.0030}}].

\bibitem{Harz:2012fz}
J.~Harz, B.~Herrmann, M.~Klasen, K.~Kovarik and Q.~L. Boulc'h,
  \emph{{Neutralino-stop coannihilation into electroweak gauge and Higgs bosons
  at one loop}},
  \href{http://dx.doi.org/10.1103/PhysRevD.87.054031}{\emph{Phys. Rev. D} {\bf
  87} (2013) 054031}, [\href{http://arxiv.org/abs/1212.5241}{{\tt 1212.5241}}].

\bibitem{Harz:2014tma}
J.~Harz, B.~Herrmann, M.~Klasen and K.~Kovarik, \emph{{One-loop corrections to
  neutralino-stop coannihilation revisited}},
  \href{http://dx.doi.org/10.1103/PhysRevD.91.034028}{\emph{Phys. Rev. D} {\bf
  91} (2015) 034028}, [\href{http://arxiv.org/abs/1409.2898}{{\tt 1409.2898}}].

\bibitem{Herrmann:2014kma}
B.~Herrmann, M.~Klasen, K.~Kovarik, M.~Meinecke and P.~Steppeler,
  \emph{{One-loop corrections to gaugino (co)annihilation into quarks in the
  MSSM}}, \href{http://dx.doi.org/10.1103/PhysRevD.89.114012}{\emph{Phys. Rev.
  D} {\bf 89} (2014) 114012}, [\href{http://arxiv.org/abs/1404.2931}{{\tt
  1404.2931}}].

\bibitem{Harz:2014gaa}
J.~Harz, B.~Herrmann, M.~Klasen, K.~Kova\v{r}\'\i{}k and M.~Meinecke,
  \emph{{SUSY-QCD corrections to stop annihilation into electroweak final
  states including Coulomb enhancement effects}},
  \href{http://dx.doi.org/10.1103/PhysRevD.91.034012}{\emph{Phys. Rev. D} {\bf
  91} (2015) 034012}, [\href{http://arxiv.org/abs/1410.8063}{{\tt 1410.8063}}].

\bibitem{Schmiemann:2019czm}
S.~Schmiemann, J.~Harz, B.~Herrmann, M.~Klasen and K.~Kova\v{r}\'\i{}k,
  \emph{{Squark-pair annihilation into quarks at next-to-leading order}},
  \href{http://dx.doi.org/10.1103/PhysRevD.99.095015}{\emph{Phys. Rev. D} {\bf
  99} (2019) 095015}, [\href{http://arxiv.org/abs/1903.10998}{{\tt
  1903.10998}}].

\bibitem{Branahl:2019yot}
J.~Branahl, J.~Harz, B.~Herrmann, M.~Klasen, K.~Kova\v{r}\'\i{}k and
  S.~Schmiemann, \emph{{SUSY-QCD corrected and Sommerfeld enhanced stau
  annihilation into heavy quarks with scheme and scale uncertainties}},
  \href{http://dx.doi.org/10.1103/PhysRevD.100.115003}{\emph{Phys. Rev. D} {\bf
  100} (2019) 115003}, [\href{http://arxiv.org/abs/1909.09527}{{\tt
  1909.09527}}].

\bibitem{Klasen:2022ptb}
M.~Klasen, K.~Kova\v{r}\'\i{}k and L.~P. Wiggering, \emph{{Radiative
  corrections to stop-antistop annihilation into gluons and light quarks}},
  \href{http://dx.doi.org/10.1103/PhysRevD.106.115032}{\emph{Phys. Rev. D} {\bf
  106} (2022) 115032}, [\href{http://arxiv.org/abs/2210.05260}{{\tt
  2210.05260}}].

\bibitem{Belanger:2016tqb}
G.~B\'elanger, V.~Bizouard, F.~Boudjema and G.~Chalons, \emph{{One-loop
  renormalization of the NMSSM in SloopS: The neutralino-chargino and sfermion
  sectors}}, \href{http://dx.doi.org/10.1103/PhysRevD.93.115031}{\emph{Phys.
  Rev. D} {\bf 93} (2016) 115031}, [\href{http://arxiv.org/abs/1602.05495}{{\tt
  1602.05495}}].

\bibitem{Belanger:2017rgu}
G.~B\'elanger, V.~Bizouard, F.~Boudjema and G.~Chalons, \emph{{One-loop
  renormalization of the NMSSM in SloopS. II. The Higgs sector}},
  \href{http://dx.doi.org/10.1103/PhysRevD.96.015040}{\emph{Phys. Rev. D} {\bf
  96} (2017) 015040}, [\href{http://arxiv.org/abs/1705.02209}{{\tt
  1705.02209}}].

\bibitem{Klasen:2013btp}
M.~Klasen, C.~E. Yaguna and J.~D. Ruiz-Alvarez, \emph{{Electroweak corrections
  to the direct detection cross section of inert higgs dark matter}},
  \href{http://dx.doi.org/10.1103/PhysRevD.87.075025}{\emph{Phys. Rev. D} {\bf
  87} (2013) 075025}, [\href{http://arxiv.org/abs/1302.1657}{{\tt 1302.1657}}].

\bibitem{Banerjee:2019luv}
S.~Banerjee, F.~Boudjema, N.~Chakrabarty, G.~Chalons and H.~Sun, \emph{{Relic
  density of dark matter in the inert doublet model beyond leading order: The
  heavy mass case}},
  \href{http://dx.doi.org/10.1103/PhysRevD.100.095024}{\emph{Phys. Rev. D} {\bf
  100} (2019) 095024}, [\href{http://arxiv.org/abs/1906.11269}{{\tt
  1906.11269}}].

\bibitem{Banerjee:2021oxc}
S.~Banerjee, F.~Boudjema, N.~Chakrabarty and H.~Sun, \emph{{Relic density of
  dark matter in the inert doublet model beyond leading order for the low mass
  region: 1. Renormalisation and constraints}},
  \href{http://dx.doi.org/10.1103/PhysRevD.104.075002}{\emph{Phys. Rev. D} {\bf
  104} (2021) 075002}, [\href{http://arxiv.org/abs/2101.02165}{{\tt
  2101.02165}}].

\bibitem{Iengo:2009ni}
R.~Iengo, \emph{{Sommerfeld enhancement: General results from field theory
  diagrams}},
  \href{http://dx.doi.org/10.1088/1126-6708/2009/05/024}{\emph{JHEP} {\bf 05}
  (2009) 024}, [\href{http://arxiv.org/abs/0902.0688}{{\tt 0902.0688}}].

\bibitem{Beneke:2016ync}
M.~Beneke, A.~Bharucha, F.~Dighera, C.~Hellmann, A.~Hryczuk, S.~Recksiegel
  et~al., \emph{{Relic density of wino-like dark matter in the MSSM}},
  \href{http://dx.doi.org/10.1007/JHEP03(2016)119}{\emph{JHEP} {\bf 03} (2016)
  119}, [\href{http://arxiv.org/abs/1601.04718}{{\tt 1601.04718}}].

\bibitem{Harz:2017dlj}
J.~Harz and K.~Petraki, \emph{{Higgs Enhancement for the Dark Matter Relic
  Density}}, \href{http://dx.doi.org/10.1103/PhysRevD.97.075041}{\emph{Phys.
  Rev. D} {\bf 97} (2018) 075041}, [\href{http://arxiv.org/abs/1711.03552}{{\tt
  1711.03552}}].

\bibitem{vonHarling:2014kha}
B.~von Harling and K.~Petraki, \emph{{Bound-state formation for thermal relic
  dark matter and unitarity}},
  \href{http://dx.doi.org/10.1088/1475-7516/2014/12/033}{\emph{JCAP} {\bf 12}
  (2014) 033}, [\href{http://arxiv.org/abs/1407.7874}{{\tt 1407.7874}}].

\bibitem{Harz:2018csl}
J.~Harz and K.~Petraki, \emph{{Radiative bound-state formation in unbroken
  perturbative non-Abelian theories and implications for dark matter}},
  \href{http://dx.doi.org/10.1007/JHEP07(2018)096}{\emph{JHEP} {\bf 07} (2018)
  096}, [\href{http://arxiv.org/abs/1805.01200}{{\tt 1805.01200}}].

\bibitem{Biondini:2018xor}
S.~Biondini, \emph{{Bound-state effects for dark matter with Higgs-like
  mediators}}, \href{http://dx.doi.org/10.1007/JHEP06(2018)104}{\emph{JHEP}
  {\bf 06} (2018) 104}, [\href{http://arxiv.org/abs/1805.00353}{{\tt
  1805.00353}}].

\bibitem{Harz:2019rro}
J.~Harz and K.~Petraki, \emph{{Higgs-mediated bound states in dark-matter
  models}}, \href{http://dx.doi.org/10.1007/JHEP04(2019)130}{\emph{JHEP} {\bf
  04} (2019) 130}, [\href{http://arxiv.org/abs/1901.10030}{{\tt 1901.10030}}].

\bibitem{Harz:2016dql}
J.~Harz, B.~Herrmann, M.~Klasen, K.~Kovarik and P.~Steppeler,
  \emph{{Theoretical uncertainty of the supersymmetric dark matter relic
  density from scheme and scale variations}},
  \href{http://dx.doi.org/10.1103/PhysRevD.93.114023}{\emph{Phys. Rev. D} {\bf
  93} (2016) 114023}, [\href{http://arxiv.org/abs/1602.08103}{{\tt
  1602.08103}}].

\bibitem{Hryczuk:2011vi}
A.~Hryczuk and R.~Iengo, \emph{{The one-loop and Sommerfeld electroweak
  corrections to the Wino dark matter annihilation}},
  \href{http://dx.doi.org/10.1007/JHEP01(2012)163}{\emph{JHEP} {\bf 01} (2012)
  163}, [\href{http://arxiv.org/abs/1111.2916}{{\tt 1111.2916}}].

\bibitem{Hryczuk:2014hpa}
A.~Hryczuk, I.~Cholis, R.~Iengo, M.~Tavakoli and P.~Ullio, \emph{{Indirect
  Detection Analysis: Wino Dark Matter Case Study}},
  \href{http://dx.doi.org/10.1088/1475-7516/2014/07/031}{\emph{JCAP} {\bf 07}
  (2014) 031}, [\href{http://arxiv.org/abs/1401.6212}{{\tt 1401.6212}}].

\bibitem{Baumgart:2017nsr}
M.~Baumgart, T.~Cohen, I.~Moult, N.~L. Rodd, T.~R. Slatyer, M.~P. Solon et~al.,
  \emph{{Resummed Photon Spectra for WIMP Annihilation}},
  \href{http://dx.doi.org/10.1007/JHEP03(2018)117}{\emph{JHEP} {\bf 03} (2018)
  117}, [\href{http://arxiv.org/abs/1712.07656}{{\tt 1712.07656}}].

\bibitem{Baumgart:2018yed}
M.~Baumgart, T.~Cohen, E.~Moulin, I.~Moult, L.~Rinchiuso, N.~L. Rodd et~al.,
  \emph{{Precision Photon Spectra for Wino Annihilation}},
  \href{http://dx.doi.org/10.1007/JHEP01(2019)036}{\emph{JHEP} {\bf 01} (2019)
  036}, [\href{http://arxiv.org/abs/1808.08956}{{\tt 1808.08956}}].

\bibitem{Drees:1992rr}
M.~Drees and M.~M. Nojiri, \emph{{New contributions to coherent neutralino -
  nucleus scattering}},
  \href{http://dx.doi.org/10.1103/PhysRevD.47.4226}{\emph{Phys. Rev. D} {\bf
  47} (1993) 4226--4232}, [\href{http://arxiv.org/abs/hep-ph/9210272}{{\tt
  hep-ph/9210272}}].

\bibitem{Drees:1993bu}
M.~Drees and M.~Nojiri, \emph{{Neutralino - nucleon scattering revisited}},
  \href{http://dx.doi.org/10.1103/PhysRevD.48.3483}{\emph{Phys. Rev. D} {\bf
  48} (1993) 3483--3501}, [\href{http://arxiv.org/abs/hep-ph/9307208}{{\tt
  hep-ph/9307208}}].

\bibitem{Hisano:2004pv}
J.~Hisano, S.~Matsumoto, M.~M. Nojiri and O.~Saito, \emph{{Direct detection of
  the Wino and Higgsino-like neutralino dark matters at one-loop level}},
  \href{http://dx.doi.org/10.1103/PhysRevD.71.015007}{\emph{Phys. Rev. D} {\bf
  71} (2005) 015007}, [\href{http://arxiv.org/abs/hep-ph/0407168}{{\tt
  hep-ph/0407168}}].

\bibitem{Berlin:2015njh}
A.~Berlin, D.~S. Robertson, M.~P. Solon and K.~M. Zurek, \emph{{Bino
  variations: Effective field theory methods for dark matter direct
  detection}}, \href{http://dx.doi.org/10.1103/PhysRevD.93.095008}{\emph{Phys.
  Rev. D} {\bf 93} (2016) 095008}, [\href{http://arxiv.org/abs/1511.05964}{{\tt
  1511.05964}}].

\bibitem{Klasen:2016qyz}
M.~Klasen, K.~Kovarik and P.~Steppeler, \emph{{SUSY-QCD corrections for direct
  detection of neutralino dark matter and correlations with relic density}},
  \href{http://dx.doi.org/10.1103/PhysRevD.94.095002}{\emph{Phys. Rev. D} {\bf
  94} (2016) 095002}, [\href{http://arxiv.org/abs/1607.06396}{{\tt
  1607.06396}}].

\bibitem{Bisal:2023fgb}
S.~Bisal, A.~Chatterjee, D.~Das and S.~A. Pasha, \emph{{Radiative Corrections
  to Aid the Direct Detection of the Higgsino-like Neutralino Dark Matter:
  Spin-Independent Interactions}},  \href{http://arxiv.org/abs/2311.09937}{{\tt
  2311.09937}}.

\bibitem{Bisal:2023iip}
S.~Bisal, A.~Chatterjee, D.~Das and S.~A. Pasha, \emph{{Confronting electroweak
  MSSM through one-loop renormalized neutralino-Higgs interactions for dark
  matter direct detection and muon $(g-2)$}},
  \href{http://arxiv.org/abs/2311.09938}{{\tt 2311.09938}}.

\bibitem{Abe:2015rja}
T.~Abe and R.~Sato, \emph{{Quantum corrections to the spin-independent cross
  section of the inert doublet dark matter}},
  \href{http://dx.doi.org/10.1007/JHEP03(2015)109}{\emph{JHEP} {\bf 03} (2015)
  109}, [\href{http://arxiv.org/abs/1501.04161}{{\tt 1501.04161}}].

\bibitem{Azevedo:2018exj}
D.~Azevedo, M.~Duch, B.~Grzadkowski, D.~Huang, M.~Iglicki and R.~Santos,
  \emph{{One-loop contribution to dark-matter-nucleon scattering in the
  pseudo-scalar dark matter model}},
  \href{http://dx.doi.org/10.1007/JHEP01(2019)138}{\emph{JHEP} {\bf 01} (2019)
  138}, [\href{http://arxiv.org/abs/1810.06105}{{\tt 1810.06105}}].

\bibitem{Ishiwata:2018sdi}
K.~Ishiwata and T.~Toma, \emph{{Probing pseudo Nambu-Goldstone boson dark
  matter at loop level}},
  \href{http://dx.doi.org/10.1007/JHEP12(2018)089}{\emph{JHEP} {\bf 12} (2018)
  089}, [\href{http://arxiv.org/abs/1810.08139}{{\tt 1810.08139}}].

\bibitem{Ghorbani:2018pjh}
K.~Ghorbani and P.~H. Ghorbani, \emph{{Leading Loop Effects in
  Pseudoscalar-Higgs Portal Dark Matter}},
  \href{http://dx.doi.org/10.1007/JHEP05(2019)096}{\emph{JHEP} {\bf 05} (2019)
  096}, [\href{http://arxiv.org/abs/1812.04092}{{\tt 1812.04092}}].

\bibitem{Abe:2018emu}
T.~Abe, M.~Fujiwara and J.~Hisano, \emph{{Loop corrections to dark matter
  direct detection in a pseudoscalar mediator dark matter model}},
  \href{http://dx.doi.org/10.1007/JHEP02(2019)028}{\emph{JHEP} {\bf 02} (2019)
  028}, [\href{http://arxiv.org/abs/1810.01039}{{\tt 1810.01039}}].

\bibitem{Glaus:2022rdc}
S.~Glaus, M.~M\"uhlleitner, J.~M\"uller, S.~Patel and R.~Santos,
  \emph{{Electroweak corrections to dark matter direct detection in the dark
  singlet phase of the N2HDM}},
  \href{http://dx.doi.org/10.1016/j.physletb.2022.137342}{\emph{Phys. Lett. B}
  {\bf 833} (2022) 137342}, [\href{http://arxiv.org/abs/2204.13145}{{\tt
  2204.13145}}].

\bibitem{Glaus:2019itb}
S.~Glaus, M.~M\"uhlleitner, J.~M\"uller, S.~Patel and R.~Santos,
  \emph{{Electroweak Corrections to Dark Matter Direct Detection in a Vector
  Dark Matter Model}},
  \href{http://dx.doi.org/10.1007/JHEP10(2019)152}{\emph{JHEP} {\bf 10} (2019)
  152}, [\href{http://arxiv.org/abs/1908.09249}{{\tt 1908.09249}}].

\bibitem{Biekotter:2022bxp}
T.~Biek\"otter, P.~Gabriel, M.~O. Olea-Romacho and R.~Santos, \emph{{Direct
  detection of pseudo-Nambu-Goldstone dark matter in a two Higgs doublet plus
  singlet extension of the SM}},
  \href{http://dx.doi.org/10.1007/JHEP10(2022)126}{\emph{JHEP} {\bf 10} (2022)
  126}, [\href{http://arxiv.org/abs/2207.04973}{{\tt 2207.04973}}].

\bibitem{Berlin:2015ymu}
A.~Berlin, D.~Hooper and S.~D. McDermott, \emph{{Dark matter elastic scattering
  through Higgs loops}},
  \href{http://dx.doi.org/10.1103/PhysRevD.92.123531}{\emph{Phys. Rev. D} {\bf
  92} (2015) 123531}, [\href{http://arxiv.org/abs/1508.05390}{{\tt
  1508.05390}}].

\bibitem{Ertas:2019dew}
F.~Ertas and F.~Kahlhoefer, \emph{{Loop-induced direct detection signatures
  from CP-violating scalar mediators}},
  \href{http://dx.doi.org/10.1007/JHEP06(2019)052}{\emph{JHEP} {\bf 06} (2019)
  052}, [\href{http://arxiv.org/abs/1902.11070}{{\tt 1902.11070}}].

\bibitem{Borschensky:2020olr}
C.~Borschensky, G.~Coniglio, B.~J\"ager, J.~Jochum and V.~Schipperges,
  \emph{{Direct detection of dark matter: Precision predictions in a simplified
  model framework}},
  \href{http://dx.doi.org/10.1140/epjc/s10052-020-08795-x}{\emph{Eur. Phys. J.
  C} {\bf 81} (2021) 44}, [\href{http://arxiv.org/abs/2008.04253}{{\tt
  2008.04253}}].

\bibitem{Haisch:2013uaa}
U.~Haisch and F.~Kahlhoefer, \emph{{On the importance of loop-induced
  spin-independent interactions for dark matter direct detection}},
  \href{http://dx.doi.org/10.1088/1475-7516/2013/04/050}{\emph{JCAP} {\bf 04}
  (2013) 050}, [\href{http://arxiv.org/abs/1302.4454}{{\tt 1302.4454}}].

\bibitem{Crivellin:2014qxa}
A.~Crivellin, F.~D'Eramo and M.~Procura, \emph{{New Constraints on Dark Matter
  Effective Theories from Standard Model Loops}},
  \href{http://dx.doi.org/10.1103/PhysRevLett.112.191304}{\emph{Phys. Rev.
  Lett.} {\bf 112} (2014) 191304}, [\href{http://arxiv.org/abs/1402.1173}{{\tt
  1402.1173}}].

\bibitem{Arbey:2008kv}
A.~Arbey and F.~Mahmoudi, \emph{{SUSY constraints from relic density: High
  sensitivity to pre-BBN expansion rate}},
  \href{http://dx.doi.org/10.1016/j.physletb.2008.09.032}{\emph{Phys. Lett. B}
  {\bf 669} (2008) 46--51}, [\href{http://arxiv.org/abs/0803.0741}{{\tt
  0803.0741}}].

\bibitem{Yaguna:2010hn}
C.~E. Yaguna, \emph{{Large contributions to dark matter annihilation from
  three-body final states}},
  \href{http://dx.doi.org/10.1103/PhysRevD.81.075024}{\emph{Phys. Rev. D} {\bf
  81} (2010) 075024}, [\href{http://arxiv.org/abs/1003.2730}{{\tt 1003.2730}}].

\bibitem{Saikawa:2020swg}
K.~Saikawa and S.~Shirai, \emph{{Precise WIMP Dark Matter Abundance and
  Standard Model Thermodynamics}},
  \href{http://dx.doi.org/10.1088/1475-7516/2020/08/011}{\emph{JCAP} {\bf 08}
  (2020) 011}, [\href{http://arxiv.org/abs/2005.03544}{{\tt 2005.03544}}].

\bibitem{Allanach:2003jw}
B.~C. Allanach, S.~Kraml and W.~Porod, \emph{{Theoretical uncertainties in
  sparticle mass predictions from computational tools}},
  \href{http://dx.doi.org/10.1088/1126-6708/2003/03/016}{\emph{JHEP} {\bf 03}
  (2003) 016}, [\href{http://arxiv.org/abs/hep-ph/0302102}{{\tt
  hep-ph/0302102}}].

\bibitem{Beneke:2014gla}
M.~Beneke, F.~Dighera and A.~Hryczuk, \emph{{Relic density computations at NLO:
  infrared finiteness and thermal correction}},
  \href{http://dx.doi.org/10.1007/JHEP10(2014)045}{\emph{JHEP} {\bf 10} (2014)
  045}, [\href{http://arxiv.org/abs/1409.3049}{{\tt 1409.3049}}].

\bibitem{Binder:2017rgn}
T.~Binder, T.~Bringmann, M.~Gustafsson and A.~Hryczuk, \emph{{Early kinetic
  decoupling of dark matter: when the standard way of calculating the thermal
  relic density fails}},
  \href{http://dx.doi.org/10.1103/PhysRevD.96.115010}{\emph{Phys. Rev. D} {\bf
  96} (2017) 115010}, [\href{http://arxiv.org/abs/1706.07433}{{\tt
  1706.07433}}].

\bibitem{Aboubrahim:2023yag}
A.~Aboubrahim, M.~Klasen and L.~P. Wiggering, \emph{{Forbidden dark matter
  annihilation into leptons with full collision terms}},
  \href{http://dx.doi.org/10.1088/1475-7516/2023/08/075}{\emph{JCAP} {\bf 08}
  (2023) 075}, [\href{http://arxiv.org/abs/2306.07753}{{\tt 2306.07753}}].

\bibitem{Hamann:2006pf}
J.~Hamann, S.~Hannestad, M.~S. Sloth and Y.~Y.~Y. Wong, \emph{{How robust are
  inflation model and dark matter constraints from cosmological data?}},
  \href{http://dx.doi.org/10.1103/PhysRevD.75.023522}{\emph{Phys. Rev. D} {\bf
  75} (2007) 023522}, [\href{http://arxiv.org/abs/astro-ph/0611582}{{\tt
  astro-ph/0611582}}].

\bibitem{Bringmann:2018jpr}
T.~Bringmann, F.~Kahlhoefer, K.~Schmidt-Hoberg and P.~Walia, \emph{{Converting
  nonrelativistic dark matter to radiation}},
  \href{http://dx.doi.org/10.1103/PhysRevD.98.023543}{\emph{Phys. Rev. D} {\bf
  98} (2018) 023543}, [\href{http://arxiv.org/abs/1803.03644}{{\tt
  1803.03644}}].

\bibitem{PhysRevLett.39.165}
B.~W. Lee and S.~Weinberg, \emph{Cosmological lower bound on heavy-neutrino
  masses}, \href{http://dx.doi.org/10.1103/PhysRevLett.39.165}{\emph{Phys. Rev.
  Lett.} {\bf 39} (Jul, 1977) 165--168}.

\bibitem{Edsjo:1997bg}
J.~Edsjo and P.~Gondolo, \emph{{Neutralino relic density including
  coannihilations}},
  \href{http://dx.doi.org/10.1103/PhysRevD.56.1879}{\emph{Phys. Rev. D} {\bf
  56} (1997) 1879--1894}, [\href{http://arxiv.org/abs/hep-ph/9704361}{{\tt
  hep-ph/9704361}}].

\bibitem{Belanger:2018ccd}
G.~B\'elanger, F.~Boudjema, A.~Goudelis, A.~Pukhov and B.~Zaldivar,
  \emph{{micrOMEGAs5.0 : Freeze-in}},
  \href{http://dx.doi.org/10.1016/j.cpc.2018.04.027}{\emph{Comput. Phys.
  Commun.} {\bf 231} (2018) 173--186},
  [\href{http://arxiv.org/abs/1801.03509}{{\tt 1801.03509}}].

\bibitem{Arbey:2009gu}
A.~Arbey and F.~Mahmoudi, \emph{{SuperIso Relic: A Program for calculating
  relic density and flavor physics observables in Supersymmetry}},
  \href{http://dx.doi.org/10.1016/j.cpc.2010.03.010}{\emph{Comput. Phys.
  Commun.} {\bf 181} (2010) 1277--1292},
  [\href{http://arxiv.org/abs/0906.0369}{{\tt 0906.0369}}].

\bibitem{Ambrogi:2018jqj}
F.~Ambrogi, C.~Arina, M.~Backovic, J.~Heisig, F.~Maltoni, L.~Mantani et~al.,
  \emph{{MadDM v.3.0: a Comprehensive Tool for Dark Matter Studies}},
  \href{http://dx.doi.org/10.1016/j.dark.2018.11.009}{\emph{Phys. Dark Univ.}
  {\bf 24} (2019) 100249}, [\href{http://arxiv.org/abs/1804.00044}{{\tt
  1804.00044}}].

\bibitem{Bringmann:2018lay}
T.~Bringmann, J.~Edsj\"o, P.~Gondolo, P.~Ullio and L.~Bergstr\"om,
  \emph{{DarkSUSY 6 : An Advanced Tool to Compute Dark Matter Properties
  Numerically}},
  \href{http://dx.doi.org/10.1088/1475-7516/2018/07/033}{\emph{JCAP} {\bf 07}
  (2018) 033}, [\href{http://arxiv.org/abs/1802.03399}{{\tt 1802.03399}}].

\bibitem{Binder:2021bmg}
T.~Binder, T.~Bringmann, M.~Gustafsson and A.~Hryczuk, \emph{{Dark matter relic
  abundance beyond kinetic equilibrium}},
  \href{http://dx.doi.org/10.1140/epjc/s10052-021-09357-5}{\emph{Eur. Phys. J.
  C} {\bf 81} (2021) 577}, [\href{http://arxiv.org/abs/2103.01944}{{\tt
  2103.01944}}].

\bibitem{Fan:2010gt}
J.~Fan, M.~Reece and L.-T. Wang, \emph{{Non-relativistic effective theory of
  dark matter direct detection}},
  \href{http://dx.doi.org/10.1088/1475-7516/2010/11/042}{\emph{JCAP} {\bf 11}
  (2010) 042}, [\href{http://arxiv.org/abs/1008.1591}{{\tt 1008.1591}}].

\bibitem{Hill:2014yka}
R.~J. Hill and M.~P. Solon, \emph{{Standard Model anatomy of WIMP dark matter
  direct detection I: weak-scale matching}},
  \href{http://dx.doi.org/10.1103/PhysRevD.91.043504}{\emph{Phys. Rev. D} {\bf
  91} (2015) 043504}, [\href{http://arxiv.org/abs/1401.3339}{{\tt 1401.3339}}].

\bibitem{Hill:2014yxa}
R.~J. Hill and M.~P. Solon, \emph{{Standard Model anatomy of WIMP dark matter
  direct detection II: QCD analysis and hadronic matrix elements}},
  \href{http://dx.doi.org/10.1103/PhysRevD.91.043505}{\emph{Phys. Rev. D} {\bf
  91} (2015) 043505}, [\href{http://arxiv.org/abs/1409.8290}{{\tt 1409.8290}}].

\bibitem{Hisano:2015bma}
J.~Hisano, R.~Nagai and N.~Nagata, \emph{{Effective Theories for Dark Matter
  Nucleon Scattering}},
  \href{http://dx.doi.org/10.1007/JHEP05(2015)037}{\emph{JHEP} {\bf 05} (2015)
  037}, [\href{http://arxiv.org/abs/1502.02244}{{\tt 1502.02244}}].

\bibitem{Shifman:1978zn}
M.~A. Shifman, A.~I. Vainshtein and V.~I. Zakharov, \emph{{Remarks on Higgs
  Boson Interactions with Nucleons}},
  \href{http://dx.doi.org/10.1016/0370-2693(78)90481-1}{\emph{Phys. Lett. B}
  {\bf 78} (1978) 443--446}.

\bibitem{Hoferichter:2015dsa}
M.~Hoferichter, J.~Ruiz~de Elvira, B.~Kubis and U.-G. Mei\ss{}ner,
  \emph{{High-Precision Determination of the Pion-Nucleon \ensuremath{\sigma}
  Term from Roy-Steiner Equations}},
  \href{http://dx.doi.org/10.1103/PhysRevLett.115.092301}{\emph{Phys. Rev.
  Lett.} {\bf 115} (2015) 092301}, [\href{http://arxiv.org/abs/1506.04142}{{\tt
  1506.04142}}].

\bibitem{Gasser:1990ce}
J.~Gasser, H.~Leutwyler and M.~E. Sainio, \emph{{Sigma term update}},
  \href{http://dx.doi.org/10.1016/0370-2693(91)91393-A}{\emph{Phys. Lett. B}
  {\bf 253} (1991) 252--259}.

\bibitem{Siegel:1979wq}
W.~Siegel, \emph{{Supersymmetric Dimensional Regularization via Dimensional
  Reduction}},
  \href{http://dx.doi.org/10.1016/0370-2693(79)90282-X}{\emph{Phys. Lett. B}
  {\bf 84} (1979) 193--196}.

\bibitem{Signer:2008va}
A.~Signer and D.~Stockinger, \emph{{Using Dimensional Reduction for Hadronic
  Collisions}},
  \href{http://dx.doi.org/10.1016/j.nuclphysb.2008.09.016}{\emph{Nucl. Phys. B}
  {\bf 808} (2009) 88--120}, [\href{http://arxiv.org/abs/0807.4424}{{\tt
  0807.4424}}].

\bibitem{Heinemeyer:2010mm}
S.~Heinemeyer, H.~Rzehak and C.~Schappacher, \emph{{Proposals for Bottom
  Quark/Squark Renormalization in the Complex MSSM}},
  \href{http://dx.doi.org/10.1103/PhysRevD.82.075010}{\emph{Phys. Rev. D} {\bf
  82} (2010) 075010}, [\href{http://arxiv.org/abs/1007.0689}{{\tt 1007.0689}}].

\bibitem{Heinemeyer:2023pcc}
S.~Heinemeyer and F.~von~der Pahlen, \emph{{Automated choice for the best
  renormalization scheme in BSM models}},
  \href{http://dx.doi.org/10.1140/epjc/s10052-023-12009-5}{\emph{Eur. Phys. J.
  C} {\bf 83} (2023) 865}, [\href{http://arxiv.org/abs/2302.12187}{{\tt
  2302.12187}}].

\bibitem{Catani:1996vz}
S.~Catani and M.~H. Seymour, \emph{{A General algorithm for calculating jet
  cross-sections in NLO QCD}},
  \href{http://dx.doi.org/10.1016/S0550-3213(96)00589-5}{\emph{Nucl. Phys. B}
  {\bf 485} (1997) 291--419}, [\href{http://arxiv.org/abs/hep-ph/9605323}{{\tt
  hep-ph/9605323}}].

\bibitem{Catani:2002hc}
S.~Catani, S.~Dittmaier, M.~H. Seymour and Z.~Trocsanyi, \emph{{The Dipole
  formalism for next-to-leading order QCD calculations with massive partons}},
  \href{http://dx.doi.org/10.1016/S0550-3213(02)00098-6}{\emph{Nucl. Phys. B}
  {\bf 627} (2002) 189--265}, [\href{http://arxiv.org/abs/hep-ph/0201036}{{\tt
  hep-ph/0201036}}].

\bibitem{Harz:2022ipe}
J.~Harz, M.~Klasen, M.~Y. Sassi and L.~P. Wiggering, \emph{{Dipole formalism
  for massive initial-state particles and its application to dark matter
  calculations}},
  \href{http://dx.doi.org/10.1103/PhysRevD.107.056020}{\emph{Phys. Rev. D} {\bf
  107} (2023) 056020}, [\href{http://arxiv.org/abs/2210.03409}{{\tt
  2210.03409}}].

\bibitem{Fabricius:1981sx}
K.~Fabricius, I.~Schmitt, G.~Kramer and G.~Schierholz, \emph{{Higher Order
  Perturbative QCD Calculation of Jet Cross-Sections in e+ e- Annihilation}},
  \href{http://dx.doi.org/10.1007/BF01578281}{\emph{Z. Phys. C} {\bf 11} (1981)
  315}.

\bibitem{Giele:1991vf}
W.~T. Giele and E.~W.~N. Glover, \emph{{Higher order corrections to jet
  cross-sections in e+ e- annihilation}},
  \href{http://dx.doi.org/10.1103/PhysRevD.46.1980}{\emph{Phys. Rev. D} {\bf
  46} (1992) 1980--2010}.

\bibitem{Harris:2001sx}
B.~W. Harris and J.~F. Owens, \emph{{The Two cutoff phase space slicing
  method}}, \href{http://dx.doi.org/10.1103/PhysRevD.65.094032}{\emph{Phys.
  Rev. D} {\bf 65} (2002) 094032},
  [\href{http://arxiv.org/abs/hep-ph/0102128}{{\tt hep-ph/0102128}}].

\bibitem{Beenakker:1996ch}
W.~Beenakker, R.~Hopker, M.~Spira and P.~M. Zerwas, \emph{{Squark and gluino
  production at hadron colliders}},
  \href{http://dx.doi.org/10.1016/S0550-3213(97)80027-2}{\emph{Nucl. Phys. B}
  {\bf 492} (1997) 51--103}, [\href{http://arxiv.org/abs/hep-ph/9610490}{{\tt
  hep-ph/9610490}}].

\bibitem{Binoth:2011xi}
T.~Binoth, D.~Goncalves~Netto, D.~Lopez-Val, K.~Mawatari, T.~Plehn and
  I.~Wigmore, \emph{{Automized Squark-Neutralino Production to Next-to-Leading
  Order}}, \href{http://dx.doi.org/10.1103/PhysRevD.84.075005}{\emph{Phys. Rev.
  D} {\bf 84} (2011) 075005}, [\href{http://arxiv.org/abs/1108.1250}{{\tt
  1108.1250}}].

\bibitem{Fuks:2013vua}
B.~Fuks, M.~Klasen, D.~R. Lamprea and M.~Rothering, \emph{{Precision
  predictions for electroweak superpartner production at hadron colliders with
  Resummino}},
  \href{http://dx.doi.org/10.1140/epjc/s10052-013-2480-0}{\emph{Eur. Phys. J.
  C} {\bf 73} (2013) 2480}, [\href{http://arxiv.org/abs/1304.0790}{{\tt
  1304.0790}}].

\bibitem{Fiaschi:2023tkq}
J.~Fiaschi, B.~Fuks, M.~Klasen and A.~Neuwirth, \emph{{Electroweak superpartner
  production at 13.6 Tev with Resummino}},
  \href{http://dx.doi.org/10.1140/epjc/s10052-023-11888-y}{\emph{Eur. Phys. J.
  C} {\bf 83} (2023) 707}, [\href{http://arxiv.org/abs/2304.11915}{{\tt
  2304.11915}}].

\bibitem{Hahn:2006nq}
T.~Hahn, \emph{{SUSY Les Houches Accord 2 I/O made easy}},
  \href{http://dx.doi.org/10.1016/j.cpc.2009.03.012}{\emph{Comput. Phys.
  Commun.} {\bf 180} (2009) 1681--1693},
  [\href{http://arxiv.org/abs/hep-ph/0605049}{{\tt hep-ph/0605049}}].

\bibitem{Hahn:1998yk}
T.~Hahn and M.~Perez-Victoria, \emph{{Automatized one loop calculations in
  four-dimensions and D-dimensions}},
  \href{http://dx.doi.org/10.1016/S0010-4655(98)00173-8}{\emph{Comput. Phys.
  Commun.} {\bf 118} (1999) 153--165},
  [\href{http://arxiv.org/abs/hep-ph/9807565}{{\tt hep-ph/9807565}}].

\bibitem{Skands:2003cj}
P.~Z. Skands et~al., \emph{{SUSY Les Houches accord: Interfacing SUSY spectrum
  calculators, decay packages, and event generators}},
  \href{http://dx.doi.org/10.1088/1126-6708/2004/07/036}{\emph{JHEP} {\bf 07}
  (2004) 036}, [\href{http://arxiv.org/abs/hep-ph/0311123}{{\tt
  hep-ph/0311123}}].

\bibitem{Allanach:2008qq}
B.~C. Allanach et~al., \emph{{SUSY Les Houches Accord 2}},
  \href{http://dx.doi.org/10.1016/j.cpc.2008.08.004}{\emph{Comput. Phys.
  Commun.} {\bf 180} (2009) 8--25}, [\href{http://arxiv.org/abs/0801.0045}{{\tt
  0801.0045}}].

\bibitem{Hahn:2004fe}
T.~Hahn, \emph{{CUBA: A Library for multidimensional numerical integration}},
  \href{http://dx.doi.org/10.1016/j.cpc.2005.01.010}{\emph{Comput. Phys.
  Commun.} {\bf 168} (2005) 78--95},
  [\href{http://arxiv.org/abs/hep-ph/0404043}{{\tt hep-ph/0404043}}].

\bibitem{Buckley:2013jua}
A.~Buckley, \emph{{PySLHA: a Pythonic interface to SUSY Les Houches Accord
  data}}, \href{http://dx.doi.org/10.1140/epjc/s10052-015-3638-8}{\emph{Eur.
  Phys. J. C} {\bf 75} (2015) 467}, [\href{http://arxiv.org/abs/1305.4194}{{\tt
  1305.4194}}].

\bibitem{ParticleDataGroup:2022pth}
{\scshape Particle Data Group} collaboration, R.~L. Workman et~al.,
  \emph{{Review of Particle Physics}},
  \href{http://dx.doi.org/10.1093/ptep/ptac097}{\emph{PTEP} {\bf 2022} (2022)
  083C01}.

\bibitem{Ellis:2022emx}
J.~Ellis, K.~A. Olive, V.~C. Spanos and I.~D. Stamou, \emph{{The CMSSM survives
  Planck, the LHC, LUX-ZEPLIN, Fermi-LAT, H.E.S.S. and IceCube}},
  \href{http://dx.doi.org/10.1140/epjc/s10052-023-11405-1}{\emph{Eur. Phys. J.
  C} {\bf 83} (2023) 246}, [\href{http://arxiv.org/abs/2210.16337}{{\tt
  2210.16337}}].

\end{thebibliography}
\end{document}